\documentclass[12pt,reqno]{amsart}

\usepackage{amsmath}
\usepackage{amsfonts}
\usepackage{amsthm}
\usepackage{amstext}

\newcommand{\dist}{{\operatorname{dist}}}
\newcommand{\Tr}{{\operatorname{Tr}}}
\newcommand{\N}{{\mathbb N}}
\newcommand{\R}{{\mathbb R}}
\newcommand{\C}{{\mathbb C}}

\theoremstyle{plain}
\newtheorem{thm}{Theorem}[section]
\newtheorem{prop}[thm]{Proposition}
\newtheorem{lemma}[thm]{Lemma}
\newtheorem{cor}[thm]{Corollary}
\theoremstyle{definition}
\newtheorem{defn}[thm]{Definition}
\newtheorem*{acknowledgement}{Acknowledgement} 

\newtheorem{remark}[thm]{Remark}{\it}{\rm}
\newtheorem{rem}[thm]{Remark}{\it}{\rm}

\numberwithin{equation}{section}

\title[Third derivative at the nucleus]{Third derivative of the
       one-electron density at the nucleus} 

\thanks{\copyright\ 2006 by the
       authors. This article may be reproduced in its entirety for
       non-commercial purposes.}

\author[S. Fournais, M. Hoffmann-Ostenhof, and T. \O. S\o rensen]
{S. Fournais \and M. Hoffmann-Ostenhof \and
T. \O stergaard S\o rensen}

\address[S. Fournais]{CNRS and Laboratoire de Math\'{e}matiques\\
         Universit\'{e} Paris-Sud - B\^{a}t 425\\
         F-91405 Orsay Cedex\\ France.}
\email{soeren.fournais@math.u-psud.fr}

\address[T. \O stergaard S\o rensen]
         {Laboratoire de Math\'{e}matiques\\
         Universit\'{e} Paris-Sud - B\^{a}t 425\\
         F-91405 Orsay Cedex\\ France.}

\address[T. \O stergaard S\o rensen (permanent address)]
        {Department of Mathematical Sciences,
         Aalborg University,
         Fredrik Bajers Vej 7G,
         DK-9220 Aalborg East, Denmark.}
\email{sorensen@math.aau.dk}

\address[M. Hoffmann-Ostenhof]
        {Fakult\"at f\"ur Mathematik,
         Universit\"at Wien,         
         Nordbergstra\ss e 15,
         A-1090 Vienna,
         Austria.} 
\email{maria.hoffmann-ostenhof@univie.ac.at}

\date{\today}

\begin{document}

\thispagestyle{empty}

\begin{abstract}
We study electron densities of eigenfunctions of ato\-mic
Schr\"o\-dinger operators. We prove the existence of
\({\widetilde\rho}{\,}'''(0)\), the third derivative of the
spherically averaged atomic density \(\widetilde\rho\) at the
nucleus. For eigenfunctions with corresponding eigenvalue below the
essential spectrum we obtain the bound
\({\widetilde\rho}{\,}'''(0)\leq-(7/12)Z^3\widetilde\rho(0)\), where
\(Z\) denotes the nuclear charge. This bound is optimal. 
\end{abstract}

\maketitle

\section{Introduction and results}
In a recent paper \cite{Ark} the present authors (together with
T. Hoff\-mann-Ostenhof (THO)) proved that electron densities of atomic
and molecular eigenfunctions are real analytic away from the positions
of the nuclei. Concerning questions of regularity of \(\rho\) it
therefore remains to study the behaviour of \(\rho\) in the vicinity
of the nuclei. A general (optimal) structure-result was obtained
recently \cite{nonisotropic}. For more detailed information, two
possible approaches are to study limits when approaching a nucleus
under a fixed angle \(\omega\in\mathbb{S}^2\), as was done in
\cite{nonisotropic}, and to study the spherical average of \(\rho\)
(here denoted \(\widetilde\rho\,\)), which is mostly interesting for
atoms. The existence of \({\widetilde\rho}{\,}'(0)\), the first
derivative of \(\widetilde\rho\) at the nucleus, and the identity
\({\widetilde\rho}{\,}'(0)=-Z\widetilde\rho(0)\) (see
\eqref{eq:KatoCusp} below) follow immediately from Kato's classical
result \cite{Kato57} on the `Cusp Condition' for the associated
eigenfunction (see also \cite{Steiner}, \cite{MHO-Seiler}). Two of
the present authors proved (with THO) the existence of
\({\widetilde\rho}{\,}''(0)\), and, for densities corresponding to
eigenvalues below the essential spectrum, a lower positive bound to
\({\widetilde\rho}{\,}''(0)\) in terms of \(\widetilde\rho(0)\) in
\cite{AHP}. In the present paper we prove the existence of
\({\widetilde\rho}{\,}'''(0)\) and derive a negative upper bound to it
(see Theorem~\ref{thm:1}). A key role in the proof is played by the
{\it a priori} estimate on \(2\)nd order derivatives of eigenfunctions
obtained in \cite{CMP2} (see also Remark~\ref{rem:higherDer} and
Appendix~\ref{app:B} below). Furthermore our investigations lead to an
improvement of the lower bound to \({\widetilde\rho}{\,}''(0)\) (see
Corollary~\ref{cor:improve}). The bounds on \(\widetilde\rho\,''(0)\)
and \(\widetilde\rho\,'''(0)\) in terms of \(\widetilde\rho(0)\) are
optimal (see Remarks~\ref{rem:Hydro1} and \ref{rem:Hydro2}).  

We turn to the precise description of the problem. We consider a
non-relativistic $N$-electron atom with a nucleus of charge \(Z\)
fixed at the origin in \(\R^3\). The Hamiltonian describing the system
is given by 
\begin{align}\label{Hbis}
   H =H_{N}(Z)=\sum_{j=1}^N\Big(-\Delta_j-
   \frac{Z}{|x_j|}\Big)
   +\sum_{1\le i<j\le N}\frac{1}{|x_i-x_j|}.
\end{align}
The positions of the $N$ electrons are denoted by ${\bf
  x}=(x_1,x_2,\dots,x_N)\in\mathbb R^{3N}$, where
$x_j=(x_{j,1},x_{j,2},x_{j,3})$ denotes the position of the $j$'th
electron in $\mathbb R^3$, and \(\Delta_j\) denotes the Laplacian with
respect to \(x_j\). For shortness, we will sometimes write 
\begin{equation}\label{H}
  H = -\Delta + V({\bf x}),
\end{equation}
where $\Delta= \sum_{j=1}^N \Delta_j$ is the $3N$-dimensional
Laplacian, and 
\begin{align}
  \label{eq:V}
  V({\bf x})=V_{N,Z}({\bf x})=\sum_{j=1}^N-
  \frac{Z}{|x_j|}
  +\sum_{1\le i<j\le N}\frac{1}{|x_i-x_j|}
\end{align}
is the complete (many-body) potential. With \(\nabla_j\) the gradient
with respect to \(x_j\), \(\nabla=(\nabla_1,\ldots,\nabla_N)\) will
denote the gradient with respect to \({\bf x}\).

It is a standard fact (see e.g.\ Kato~\cite{Kato51}) that $H$ is
selfadjoint with operator domain $\mathcal{D}(H)=W^{2,2}(\mathbb
R^{3N})$ and quadratic form domain \(\mathcal{Q}(H)=W^{1,2}(\mathbb
R^{3N})\).  

We consider eigenfunctions $\psi$ of $H$, i.e., solutions $\psi \in
L^2(\mathbb R^{3N})$ to the equation 
\begin{equation}\label{Hpsi}
  H\psi=E\psi,
\end{equation}
with $E\in {\mathbb R}$. To simplify notation we assume from now on,
without loss, that \(\psi\) is real. Apart from the wave function
$\psi$ itself, the most important quantity describing the state of the
atom is the {\bf one-electron density} $\rho$. It is defined by 
\begin{align}\label{rhohat}
  \rho(x)=&\sum_{j=1}^N\rho_j(x)=
  \sum_{j=1}^N
  \int_{\mathbb R^{3N-3}}|\psi(x,\hat{\bf{x}}_j)|^2
  d\hat{\bf{x}}_j\,,
\end{align}
where we use the notation 
\begin{align}
  \hat{\bf{x}}_j&=(x_1,\dots,x_{j-1}, x_{j+1},\dots, x_N)
  \intertext{ and }
  d\hat{\bf{x}}_j&=
  dx_1\dots dx_{j-1} dx_{j+1}\dots dx_N
\end{align} 
and, by abuse of notation, identify \((x_1,\dots,x_{j-1},x,
x_{j+1},\dots, x_N)$ and  \((x,\hat{\bf x}_j)\).

We assume throughout when studying \(\rho\) that \(E\) and \(\psi\) in
\eqref{Hpsi} are such that there exist constants \(C_0,\gamma>0\) such
that 
\begin{align}
  \label{eq:exp-dec}
  |\psi({\bf x})|\leq C_0\, e^{-\gamma|{\bf x}|}\quad\text{ for all } {\bf
    x}\in\mathbb R^{3N}.
\end{align}
The {\it a priori} estimate \cite[Theorem 1.2]{AHP} (see also
\cite[Remark 1.7]{AHP}) and \eqref{eq:exp-dec} imply the existence of
constants \(C_{1}, \gamma_{1}>0\) such that
\begin{align}
  \label{eq:dec_grad_psi}
  \big|\nabla\psi(\mathbf x)\big|\leq C_{1}\,e^{-\gamma_{1}|\mathbf x|}
   \ \text{ for almost all } \mathbf x\in\mathbb R^{3N}.
\end{align}
\begin{remark}
\label{rem:exp-dec}
Since \(\psi\) is continuous (see Kato~\cite{Kato57}),
\eqref{eq:exp-dec} is only an assumption on the behaviour at infinity.
For references on the exponential decay of eigenfunctions, see e.g.\
Froese and Herbst~\cite{FroeseHerbst} and Simon~\cite{Si-semi}. The
proofs of our results rely (if not indicated otherwise) on some kind
of decay-rate for \(\psi\); exponential decay is not essential, but
assumed for convenience. Note that \eqref{eq:exp-dec} and
\eqref{eq:dec_grad_psi} imply that \(\rho\) is Lipschitz continuous in
\(\R^3\) by Lebesgue's theorem on dominated convergence.
\end{remark}
In \cite{Ark} we proved (with THO) that \(\rho\) is real analytic away
from the position of the nucleus. More generally, for a molecule with
\(K\) fixed nuclei at \(R_1,\ldots,R_K\), \(R_j\in\R^3\), it was
proved that \(\rho\in C^{\omega}(\mathbb
R^3\setminus\{R_1,\ldots,R_K\})\); see also \cite{CMP1} and
\cite{taxco}. Note that the proof of analyticity does {\it not}
require any decay of \(\rho\) (apart from \(\psi\in W^{2,2}(\mathbb
R^{3N})\)). That \(\rho\) itself is not analytic at the positions of
the nuclei is already clear for the groundstate of `Hydrogenic atoms'
(\(N=1\)); in this case, \(\nabla\rho\) is not even continuous at
\(x=0\). However, as was proved in \cite{nonisotropic},
\(e^{Z|x|}\rho\in C^{1,1}(\R^3)\).

To obtain more information about the behaviour of the density at the
positions of the nuclei one therefore has to study the regularity of
other quantities, derived from \(\rho\).

One possibility is to study the function
\(r\mapsto\rho(r,\omega):=\rho(r\omega)\) for fixed
\(\omega=\frac{x}{|x|}\in\mathbb{S}^2\) (\(r=|x|\)); results in this
direction were derived by the authors (with THO) in \cite[Theorem
1.5]{nonisotropic}. In particular, for the case of atoms it was proved
that for all \(\omega\in\mathbb{S}^2\), 
\begin{align}
  \label{eq:regFixedAngle}
  \rho(\cdot,\omega)\in C^{2,\alpha}([0,\infty))\ \text{ for all }
  \alpha\in(0,1)\,,
\end{align}
and the 1st and 2nd radial derivatives were investigated at \(r=0\). 

The main quantity studied in this paper is the spherical average of
\(\rho\), 
\begin{align}
  \label{def:tildeRho}
  \widetilde\rho(r)=\int_{\mathbb S^2}\rho(r\omega)\,d\omega\ , \quad r\in[0,\infty).
\end{align}
It follows from the analyticity of \(\rho\) mentioned above that also
\(\widetilde\rho\in C^{\omega}((0,\infty))\), and from the Lipschitz
continuity of \(\rho\) in \(\R^3\) that \(\widetilde\rho\) is
Lipschitz continuous in \([0,\infty)\). 

The existence of \(\widetilde\rho\,'(0)\), the continuity of
\(\widetilde\rho\,'\) at \(r=0\), and the Cusp Condition
\begin{align}
  \label{eq:KatoCusp}
  \widetilde\rho\,'(0)=-Z\widetilde\rho(0), 
\end{align}
follows from a similar result for \(\psi\) itself by Kato
\cite{Kato57}; see \cite{Steiner}, \cite{MHO-Seiler}, and \cite[Remark
1.13]{AHP}.

To investigate properties of \(\rho\) and the derived quantities above
it is essential that \(\rho\) satisfies a differential equation. Such
an equation easily follows via \eqref{Hpsi} from 
\begin{equation}\label{hatint}
  \sum_{j=1}^N\int_{\mathbb R^{3N-3}}\psi(x, \mathbf{\hat
    x}_j)(H-E)\psi(x,\mathbf{\hat x}_j)\,d 
  \mathbf{\hat x}_j=0\,.
\end{equation} 
This implies that \(\rho\) satisfies, in the distributional sense, the
(inhomogeneous one-particle Schr\"odinger) equation
\begin{align}
   \label{eq:firstRho}
   {}-\frac12\Delta\rho-\frac{Z}{|x|}\rho+h=0\quad \text{in}\quad
   \mathbb R^3\,.
\end{align}
The function \(h\) in \eqref{eq:firstRho} is given by
\begin{align}
  \label{def:h}
  h(x)&=\sum_{j=1}^{N} h_j(x),\\
  \label{def:h-j}
  h_j(x)&=
  \int_{\R^{3N-3}}|\nabla\psi(x,\hat{\bf x}_j)|^{2}\,d\hat{\bf x}_j
  -\sum_{\ell=1,\ell\neq j}^{N}
  \int_{\R^{3N-3}}\frac{Z}{|x_{\ell}|}\,|\psi(x,\hat{\bf
    x}_j)|^{2}\,d\hat{\bf x}_j\nonumber\\ 
  &{}\quad+
  \sum_{\ell=1,\ell\neq j}^{N}
  \int_{\R^{3N-3}}\frac{1}{|x-x_{\ell}|}\,|\psi(x, \hat{\bf
    x}_j)|^{2}\,d\hat{\bf x}_j\\
  &{}\quad+
  \sum_{1\leq k<\ell\leq N,\,k\neq j\neq \ell}
  \int_{\R^{3N-3}}\frac{1}{|x_{k}-x_{\ell}|}\,|\psi(x, \hat{\bf
    x}_j)|^{2}\,d\hat{\bf x}_j 
  - E\rho_j(x)\nonumber.
\end{align}
The equation \eqref{eq:firstRho} implies that the function
\(\widetilde\rho\) in \eqref{def:tildeRho} satisfies 
\begin{align}
   \label{eq:secondRho}
   {}-\frac12\Delta\widetilde\rho-\frac{Z}{r}\widetilde\rho
   +\widetilde h=0 \ \ \text{ for }\ \ r\in(0,\infty)\,,
\end{align}
where
\(\Delta=\frac{d^2}{dr^2}+\frac{2}{r}\frac{d}{dr}
=\frac{1}{r^2}\frac{d}{dr}(r^2\frac{d}{dr})\),
and  
\begin{align}
  \label{def:hTilde}
  \widetilde h(r)=\int_{\mathbb{S}^2}h(r\omega)\,d\omega \,.
\end{align}

In \cite[Theorem 1.11]{AHP} information about the regularity of
\(\psi\) was used to prove that \(\widetilde h\in C^0([0,\infty))\),
and, using \eqref{eq:secondRho}, that consequently,
\(\widetilde\rho\in C^2([0,\infty))\), and 
\begin{align}
  \label{eq:tildeRho2}
  \widetilde\rho\,''(0)
  =\frac23\big(\,\widetilde h(0)+Z^2\widetilde\rho(0)\big)\,.
\end{align}
Moreover, denote by \(\sigma(H_N(Z))\) the spectrum of \(H_N(Z)\), and
define  
\begin{align}
  \label{eq:ionEnergy}
  \varepsilon:=E^0_{N-1}(Z)-E\ , \quad
   E^0_{N-1}(Z)=\inf\sigma(H_{N-1}(Z))\,.
\end{align}
Then if \(\varepsilon\ge0\) \cite[Theorem 1.11]{AHP}, we have
\begin{align}
  \label{eq:ionBoundH}
  h(x)\geq \varepsilon\rho(x)\ \text{ for all } x\in\R^{3}\,,
\end{align}
and so in this case, \eqref{eq:tildeRho2} implies that
\begin{align}
  \label{eq:secondRhoSign}
   \widetilde\rho\,''(0)\geq
   \frac{2}{3}\big(Z^2+\varepsilon\big)\widetilde\rho(0)
   \ge\frac{2}{3}\,Z^2\widetilde\rho(0)\,.  
\end{align}

Our main result in this paper is the following. 
\begin{thm} \label{thm:1}
Let $\psi \in L^2({\mathbb R}^{3N})$ be an atomic  eigenfunction,
\(H_N(Z)\psi\) \(= E\psi\), satisfying \eqref{eq:exp-dec}, with
associated spherically averaged density $\widetilde\rho$ defined by
\eqref{rhohat} and \eqref{def:tildeRho}. Let \(\widetilde h\) be
defined by \eqref{def:h}--\eqref{def:h-j}, \eqref{def:hTilde}, and let
\(\varepsilon\) be given by \eqref{eq:ionEnergy}. Let finally
\(\varphi_j(x,\hat{\bf x}_j)=e^{\frac{Z}{2}{|x|}}\psi(x,\hat{\bf
  x}_j)\), \(j=1,\ldots,N\).

Then \(\widetilde\rho\in C^{3}([0,\infty))\), and 
\begin{align}
  \label{eq:main}
  \widetilde\rho\,'''(0)
  &=\widetilde h\,'(0)
  -\frac{Z}{3}\big[\,\widetilde h(0)+Z^{2}\widetilde\rho(0)\big]\,
  \\
  &={}-\frac{7}{12}Z^3\widetilde\rho(0)
  -4\pi Z\sum_{j=1}^{N}\Big[\int_{\R^{3N-3}}
  |\nabla_j\varphi_j(0,\hat{\bf x}_j)|^2
  \,d\hat{\bf x}_j
  \label{eq:mainBIS1}
  \\&\qquad\qquad\quad{}+\frac{5}{3}\langle\psi(0,\cdot),
  [H_{N-1}(Z-1)-E] \psi(0,\cdot)
  \rangle_{L^2( \R^{3N-3}_{\hat{\bf x}_j})}\Big]\,.
  \nonumber
\end{align}
If \(\varepsilon\ge0\), then
\begin{align}
  \label{eq:mainBIS2}
  \widetilde\rho\,'''(0)\le{}-\frac{Z}{12}\big(7Z^2
  +20\varepsilon\big)\,\widetilde\rho(0)
  \le{}-\frac{7}{12}Z^3\widetilde\rho(0)\,.
\end{align}
\end{thm}
\begin{rem}\label{rem:higherDer}
The existence of \(\widetilde\rho\,^{(k)}(0\)) for all $k>3$ remains
an open problem.  The two main steps in the proof of
Theorem~\ref{thm:1} are Propositions~\ref{lem:h'} and
\ref{prop:allCusps} below. From the latter one sees that the existence
of \(\widetilde h\,^{(k-2)}(0)\) is necessary to prove  existence of
\(\widetilde\rho\,^{(k)}(0)\). In Proposition~\ref{lem:h'} the
existence of \(\widetilde h\,'(0)\) is proved and this result already
heavily relies on the optimal regularity results for $\psi$ (involving
an {\it a priori} estimate for second order partial derivatives of
\(\psi\)) obtained in \cite{CMP2} (see also Appendix~\ref{app:B}
below). 
\end{rem}
\begin{rem}\label{rem:sym} Note that the inequality
\begin{align}
  \label{eq:mainBIS3}
  \widetilde\rho\,'''(0)
  \le{}-\frac{7}{12}Z^3\widetilde\rho(0)
\end{align}
follows from \eqref{eq:mainBIS1} as soon as \(\psi\) is such that
\begin{align}
  \label{eq:exptValue}
  \sum_{j=1}^{N}\,\langle\psi(0,\cdot),
  [H_{N-1}(Z-1)-E] \psi(0,\cdot)
  \rangle_{L^2( \R^{3N-3}_{\hat{\bf x}_j})}\ge0\,.
\end{align}
There are cases where \eqref{eq:exptValue} holds even if the
assumption \(\varepsilon\ge0\) does not. For instance when \(E\) is an
embedded eigenvalue for the full operator \(H\) (\(\varepsilon<0\)),
but non-imbedded for the operator restricted to a symmetry
subspace. In particular, \eqref{eq:mainBIS3} holds for the fermionic
ground state. 
\end{rem}
\begin{rem}\label{rem:Hydro1}
Compare \eqref{eq:KatoCusp}, \eqref{eq:secondRhoSign}, and
Theorem~\ref{thm:1} with the fact that for the ground state of
`Hydrogenic atoms' (\(N=1\)), the corresponding density
\(\widetilde\rho_1(r)=c\,e^{-Zr}\) satisfies
\begin{align}
  \label{eq:cuspsHydro}
  \widetilde{\rho}_1{}\!^{(k)}(0)=(-Z)^k\widetilde\rho_1(0)\,.
\end{align}
In fact, if \((-\Delta-Z/|x|)\psi_n=E_n\psi_n\), \(E_n=-Z^2/4n^2\),
\(n\in\N\), \(\psi_n(x)=e^{-\frac{Z}{2}|x|}\phi_n(x)\), then
\eqref{eq:mainBIS1} implies that the corresponding density
\(\widetilde\rho_n\) satisfies  
\begin{align}
  \label{eq:cuspsGenHydro}   \nonumber
  \widetilde{\rho}_n{}\!\!'''(0)&=\big[-\frac{7}{12}Z^3
  +\frac{5}{3}ZE\big]\widetilde\rho_n(0)  
  -4\pi Z|\nabla\phi_n(0)|^2
  &\\&={}-\frac{Z^3}{12}\big[7+\frac{5}{n^2}\big]\widetilde\rho_n(0)
  -4\pi Z|\nabla\phi_n(0)|^2\,.
\end{align}
For the ground state, i.e., for \(n=1\), \(E_1=-Z^2/4\),
\(\phi_1\equiv1\), this reduces to \eqref{eq:cuspsHydro} with
\(k=3\). 

Furthermore, for {\it s} - states (zero angular momentum), we get that
\(\nabla\phi_n(0)=0\), since \(\phi_n\) is radial and \(C^{1,\alpha}\)
(see \eqref{eq:firstPhi1} below). Taking \(n\) large in
\eqref{eq:cuspsGenHydro} illustrates the quality of the bound
\eqref{eq:mainBIS2}.
\end{rem}
The proof of Theorem~\ref{thm:1} is based on the following result on
\(h\). Its proof is given in Section~\ref{hprime}.  
\begin{prop}\label{lem:h'}
Let $\psi \in L^2({\mathbb R}^{3N})$ be an atomic eigenfunction,\newline
\(H_N(Z)\psi=E\psi\), satisfying \eqref{eq:exp-dec}, and let \(h\) be
as defined in \eqref{def:h}--\eqref{def:h-j}. Let
\(\omega\in\mathbb{S}^2\) and \(\widetilde
h(r)=\int_{\mathbb{S}^2}h(r\omega)\,d\omega\).

Then both \(\widetilde h\) and the function \(r\mapsto
h(r,\omega):=h(r\omega)\) belong to\\\noindent \(C^{1}([0,\infty))\).  

Furthermore, with \(\varphi_j(x,\hat{\bf
  x}_j)=e^{\frac{Z}{2}{|x|}}\psi(x,\hat{\bf x}_j)\), \(j=1,\ldots,N\), 
\begin{align}
  \label{eq:h(0)}
  \widetilde h(0)&=
  \frac{Z^2}{4}\widetilde\rho(0)
  +4\pi\sum_{j=1}^{N}\Big[\int_{\R^{3N-3}}
  |\nabla_j\varphi_j(0,\hat{\bf x}_j)|^2
  \,d\hat{\bf x}_j
  \\&\quad\qquad{}+\langle\psi(0,\cdot),
  [H_{N-1}(Z-1)-E] \psi(0,\cdot)
  \rangle_{L^2( \R^{3N-3}_{\hat{\bf x}_j})}\Big]\,,
  \displaybreak[0]  
  \nonumber\\
  \label{eq:h'(0)}\nonumber
  \widetilde h\,'(0)&={}-Z\widetilde h(0)
  +\frac{Z^3}{12}\widetilde\rho(0)
  +\frac{4\pi}{3}Z\sum_{j=1}^{N}\Big[\int_{\R^{3N-3}}
  |\nabla_j\varphi_j(0,\hat{\bf x}_j)|^2
  \,d\hat{\bf x}_j
  \\&\qquad\quad{}-\langle\psi(0,\cdot),
  [H_{N-1}(Z-1)-E] \psi(0,\cdot)
  \rangle_{L^2( \R^{3N-3}_{\hat{\bf x}_j})}\Big]\,.
\end{align}
\end{prop}
As a byproduct of \eqref{eq:h(0)} we get the following improvement of
\eqref{eq:secondRhoSign}.
\begin{cor}\label{cor:improve}
Let $\psi \in L^2({\mathbb R}^{3N})$ be an atomic  eigenfunction,
\(H_N(Z)\psi\) \(= E\psi\), satisfying \eqref{eq:exp-dec}, with
associated spherically averaged density $\widetilde\rho$ defined by
\eqref{rhohat} and \eqref{def:tildeRho}. Let \(\varepsilon\) be given
by \eqref{eq:ionEnergy}, and assume \(\varepsilon\ge0\).

Then
\begin{align}
  \label{eq:secondRhoSignImproved}
  \widetilde\rho\,''(0)\geq
  \frac{2}{3}\big[\frac{5Z^2}{4}+\varepsilon\big]\widetilde\rho(0)
  \geq\frac{5}{6}\,Z^2\widetilde\rho(0) \,.
\end{align}
\end{cor}
\begin{proof}
Using the HVZ-theorem \cite[Theorem XIII.17]{R&S4}, \eqref{eq:h(0)}
provides an improvement of the bound \eqref{eq:ionBoundH} for \(r=0\) to
\begin{align}
  \label{eq:hImproved}
  \widetilde h(0)\ge
  \big[\frac{Z^2}{4}+\varepsilon\big]\widetilde\rho(0) \,.
\end{align}
This, using \eqref{eq:tildeRho2}, gives
\eqref{eq:secondRhoSignImproved}. 
\end{proof}
\begin{rem}\label{rem:Hydro2}
For `Hydrogenic atoms' (\(N=1\)) (see Remark~\ref{rem:Hydro1}),
\eqref{eq:tildeRho2} and \eqref{eq:h(0)} imply
\begin{align}
  \label{eq:secondHydro}
  \widetilde{\rho}_n{}\!\!''(0)&=
  \frac{Z^2}{6}\big[5+\frac{1}{n^2}\big]\widetilde\rho_n(0)
  +\frac{8\pi}{3} |\nabla\phi_n(0)|^2\,,
\end{align}
which illustrates the quality of the bound
\eqref{eq:secondRhoSignImproved} above (see also the discussion in
Remark~\ref{rem:Hydro1}), and reduces to \eqref{eq:cuspsHydro} with
\(k=2\) for the ground state (\(n=1, \phi_1\equiv1\)). 
\end{rem}
We outline the structure of the rest of the paper. In
Section~\ref{proof}, we use Proposition~\ref{lem:h'} and the equation
for \(\widetilde\rho\) (see \eqref{eq:secondRho}) to prove
Theorem~\ref{thm:1}. In Section~\ref{studyH} we then prove
Proposition~\ref{lem:h'}. This is done applying the characterization
of the regularity of the eigenfunction \(\psi\) up to order
\(C^{1,1}\) proved in \cite{CMP2} (see also Appendix~\ref{app:B} and
Lemma~\ref{lem:phi}) to the different terms in
\eqref{def:h}--\eqref{def:h-j}.  
\section{Proof of Theorem~\ref{thm:1}}\label{proof}
That \(\widetilde\rho\in C^3([0,\infty))\) and the formula
\eqref{eq:main} follow from Proposition~\ref{prop:allCusps} below
(with \(k=1\)), using Proposition~\ref{lem:h'} and
\eqref{eq:tildeRho2}. The formula \eqref{eq:mainBIS1} then follows
from \eqref{eq:main} and Proposition~\ref{lem:h'}.

If \(\varepsilon\ge0\), then the HVZ-theorem \cite[Theorem
XIII.17]{R&S4} implies that 
\begin{align}
  \label{eq:useHVZ}
  \sum_{j=1}^N\langle\psi(0,\cdot),
  [H_{N-1}(Z-1)&-E] \psi(0,\cdot)
  \rangle_{L^2( \R^{3N-3}_{\hat{\bf x}_j})}
  \\
  &\ge\varepsilon\sum_{j=1}^N\langle\psi(0,\cdot),
  \psi(0,\cdot)\rangle_{L^2(\R^{3N-3}_{\hat{\bf x}_j})} 
  =\varepsilon\rho(0)\,,
  \nonumber
\end{align}
which, together with \eqref{eq:mainBIS1}, implies \eqref{eq:mainBIS2},
since \(\widetilde\rho(0)=4\pi\rho(0)\). 

It therefore remains to prove the following proposition.
\begin{prop}
  \label{prop:allCusps}
Let $\psi \in L^2({\mathbb R}^{3N})$ be an atomic  eigenfunction,\newline
\(H_N(Z)\psi= E\psi\), satisfying \eqref{eq:exp-dec}, with associated
spherically averaged density $\widetilde\rho$ defined by
\eqref{rhohat} and \eqref{def:tildeRho}, and let \(\widetilde h\) be
as defined in \eqref{def:h}--\eqref{def:h-j} and
\eqref{def:hTilde}. Let \(k\in\N\cup\{0\}\).
 
If \(\widetilde h\in C^k([0,\infty))\) then \(\widetilde\rho\in
C^{k+2}([0,\infty))\), and
\begin{align}
  \label{eq:genCusps}
  \widetilde\rho\,^{(k+2)}(0)
  =\frac{2}{k+3}\big[(k+1)\widetilde
  h^{(k)}(0)-Z\widetilde\rho\,^{(k+1)}(0)\big]\,. 
\end{align}
\end{prop}
\begin{proof}
Let \(r>0\); multiplying \eqref{eq:secondRho} with \(r^2\),
integrating over \([\delta,r]\) for \(0<\delta<r\), and then taking
the limit \(\delta\downarrow0\), using that \(\widetilde h,
\widetilde\rho\), and \(\widetilde\rho\,'\) are all continuous on
\([0,\infty)\) (see the introduction), it follows from Lebesgue's
theorem on dominated convergence that
\begin{align}
  \label{eq:1}\nonumber
  {\widetilde\rho}\,'(r)&=\frac{2}{r^2}\int_0^r
  \big[-Z\widetilde\rho(s)s+\widetilde h(s)s^2\big]\,ds 
  \\&=-2Z\int_0^1\widetilde\rho(r\sigma)\sigma\,d\sigma
  +\frac{2}{r^2}\int_0^r\widetilde h(s)s^2\,ds\,.
\end{align}
Using again Lebesgue's theorem on dominated convergence (in the form
of Proposition~\ref{prop:genLeb} below) we get that for \(r>0\),
\begin{align}
  \label{eq:2}
  \widetilde\rho\,''(r)=2\big[\,\widetilde
  h(r)-\int_0^1\big[Z\widetilde\rho\,'(r\sigma) 
  +2\widetilde h(r\sigma)\big]\sigma^2\,d\sigma\big]\,.
\end{align}
Since \(\widetilde\rho\in C^2([0,\infty))\) (see the introduction),
\eqref{eq:2} extends to \(r=0\) by continuity and Lebesgue's
theorem. This finishes the proof in the case \(k=0\). 

For \(k\in\N\), applying Lebegue's theorem to \eqref{eq:2} it is easy
to prove by induction that if \(\widetilde h\in C^k([0,\infty))\) then
\(\widetilde\rho\in C^{k+2}([0,\infty))\), and that for \(r\ge0\), 
\begin{align}
  \label{eq:3}
  \widetilde\rho\,^{(k+2)}(r)=2\big[\,\widetilde h^{(k)}(r)
  -\int_0^1\big[Z\widetilde\rho\,^{(k+1)}(r\sigma)
  +2\widetilde h^{(k)}(r\sigma)\big]\sigma^{k+2}\,d\sigma\,\big]\,.
\end{align}
In particular, \eqref{eq:genCusps} holds. This finishes the proof of
the proposition.
\end{proof}
This finishes the proof of Theorem~\ref{thm:1}.
\qed

It remains to prove Proposition~\ref{lem:h'}.
\section{Study of the function $h$}\label{studyH}
\subsection{Proof of Proposition~\ref{lem:h'}}\label{hprime}
It clearly suffices to prove the statements in Proposition~\ref{lem:h'} for
each \(h_j\) (\(j=1,\ldots,N\)) in
\eqref{def:h-j}. Proposition~\ref{lem:h'} then follows for \(h\) by
summation. We shall prove the statements in Proposition~\ref{lem:h'}
for \(h_1\); the proof for the other \(h_j\) is completely analogous. 

Recall (see \eqref{def:h}--\eqref{def:h-j}) that \(h_1\) is defined by
\begin{align}
  \label{def:hBis}
  h_1(x)&=t_1(x)-v_1(x)+w_1(x)-E\rho_1(x),\\
  \label{def:t}
  t_1(x)&=
  \int_{\R^{3N-3}}|\nabla\psi(x,\hat{\bf x}_1)|^{2}\,d\hat{\bf
    x}_1\,,\\ 
  \label{def:v}
  v_1(x)&=
  \sum_{k=2}^{N}\int_{\R^{3N-3}}\frac{Z}{|x_{k}|}\,|\psi(x,\hat{\bf  
    x}_1)|^{2}\,d\hat{\bf x}_1\,,\\
  \label{def:w}
  w_1(x)&=\sum_{k=2}^{N}
  \int_{\R^{3N-3}}\frac{1}{|x-x_{k}|}\,|\psi(x, \hat{\bf 
    x}_1)|^{2}\,d\hat{\bf x}_1
  \nonumber\\ &{}\quad+
  \sum_{2\leq k<\ell\leq N}
  \int_{\R^{3N-3}}\frac{1}{|x_{k}-x_{\ell}|}\,|\psi(x, \hat{\bf
    x}_1)|^{2}\,d\hat{\bf x}_1\,,\\ 
  \label{def:rho1}
  \rho_1(x)&=\int_{\R^{3N-3}}|\psi(x, \hat{\bf 
    x}_1)|^{2}\,d\hat{\bf x}_1.
\end{align}
Here, \( \hat{{\bf x}}_1=(x_2, \ldots, x_N)\in\R^{3N-3}\) and
$d\hat{\bf{x}}_1= dx_2\dots dx_N$. 

We shall look at the different terms in
\eqref{def:t}--\eqref{def:rho1} separately. The statements on
regularity in Proposition~\ref{lem:h'} clearly follow from
Proposition~\ref{prop:cusps} and Proposition~\ref{prop:t'} below and
the regularity properties of \(\rho\), see the discussion in the
introduction (in the vicinity of
\eqref{eq:regFixedAngle}--\eqref{eq:KatoCusp}; see also \cite[Theorem
1.5]{nonisotropic} and \cite[Theorem 1.11]{AHP}). The formulae
\eqref{eq:h(0)}--\eqref{eq:h'(0)} follow from the formulae in
Proposition~\ref{prop:cusps} and Proposition~\ref{prop:t'}, using
\eqref{def:hBis}--\eqref{def:rho1}, and the fact that
\begin{align}
   \label{eq:expValueA}\nonumber
   \int_{\R^{3N-3}}
   &\Big\{|\nabla_{\hat{\bf x}_1} 
           \psi(0,\hat{\bf x}_1)|^2
          +\big(V_{N-1,Z-1}(\hat{\bf x}_1)-E\big)
          |\psi(0,\hat{\bf x}_1)|^2
   \Big\}\,   
   \,d\hat{\bf x}_1
   \\&= \langle\psi(0,\cdot),
   [H_{N-1}(Z-1)-E] \psi(0,\cdot)
   \rangle_{L^2( \R^{3N-3}_{\hat{\bf x}_1})}\,.
\end{align}
For the definitions of \(V_{N-1,Z-1}\) and \(H_{N-1}(Z-1)\), see
\eqref{eq:V} and \eqref{Hbis}, respectively.
\begin{prop}\label{prop:cusps}   
Let \(v_1\) and \(w_1\) be defined as in \eqref{def:v}--\eqref{def:w}
and let \(\omega\in\mathbb{S}^2\). Define \(\widetilde
v_1(r)=\int_{\mathbb{S}^2}v_1(r\omega)\,d\omega\), \(\widetilde
w_1(r)=\int_{\mathbb{S}^2}w_1(r\omega)\,d\omega\). 

Then \(v_1(\,\cdot\,,\omega), w_1(\,\cdot\,,\omega),
\widetilde{v}_1\), and \(\widetilde{w}_1\) all belong to
\(C^1\big([0,\infty)\big)\), and
\begin{align}
  \label{eq:cuspV}
  \widetilde{v}_1{}\!'(0)&={}-Z\widetilde{v}_1(0),\\ 
  \label{eq:cuspW}
  \widetilde{w}_1{}\!'(0)&={}-Z\widetilde{w}_1(0).
\end{align}
\end{prop}
\begin{rem}
Similar statements hold for \(v_j\) and \(w_j\) (\(j=2,\ldots,N\)),
and therefore, by summation, for \(v=\sum_{j=1}^Nv_j\),
\(w=\sum_{j=1}^Nw_j\). Compare this with \eqref{eq:KatoCusp}:
\({\widetilde{\rho}}{\,}'(0)=-Z\widetilde\rho(0)\); that is, three of
the four terms in \(\widetilde h\) (see \eqref{def:h}) satisfy the
`Cusp Condition' \({\widetilde{f}}{\,}'(0)=-Z\widetilde{f}(0)\). 
\end{rem}
For the remaining term in \eqref{def:hBis}, we have the following.
\begin{prop}\label{prop:t'}
Let \(t_1\) be defined as in \eqref{def:t} and let
\(\omega\in\mathbb{S}^2\). Define
\(\widetilde{t}_1(r)=\int_{\mathbb{S}^2}t_1(r\omega)\,d\omega\).

Then both \(\widetilde{t}_1\) and the function \(r\mapsto
t_1(r,\omega):=t_1(r\omega)\) belong to
\(C^1\big([0,\infty)\big)\). Furthermore, with \(\varphi_1(x,\hat{\bf
  x}_1)=e^{\frac{Z}{2}{|x|}}\psi(x,\hat{\bf x}_1)\),
\begin{align}
  \label{eq:t(0)}
  \widetilde t_1(0)&=  
  \frac{Z^2}{4}\widetilde\rho_1(0)
  +4\pi\Big[\int_{\R^{3N-3}}|\nabla_1\varphi_1(0,\hat{\bf x}_1)|^2
  \,d\hat{\bf x}_1
  \\&\qquad\qquad\qquad\qquad\qquad
  +\int_{\R^{3N-3}}|\nabla_{\hat{\bf
      x}_1}\psi(0,\hat{\bf x}_1)|^2 
  \,d\hat{\bf x}_1\Big]\,,
  \nonumber\\
  \label{eq:tCusp}
  \widetilde t_1{}\!'(0)&=-Z\widetilde t_1(0)+  
  \frac{Z^3}{12}\widetilde\rho_1(0)
  +\frac{4\pi}{3}Z\Big[\int_{\R^{3N-3}}
      |\nabla_1\varphi_1(0,\hat{\bf x}_1)|^2
  \,d\hat{\bf x}_1
  \\&\!\!\!\!\!\!\!\!\!\!\!\!\!\!
  -\int_{\R^{3N-3}}
  \Big\{\,|\nabla_{\hat{\bf x}_1}\psi(0,\hat{\bf x}_1)|^2
  +\big(V_{N-1,Z-1}(\hat{\bf x}_1)-E\big)|\psi(0,\hat{\bf
     x}_1)|^2\Big\}
  \,d\hat{\bf x}_1\Big]\,,
  \nonumber
\end{align}
with \(V_{N-1,Z-1}\) given by \eqref{eq:V}.
\end{prop}
\begin{rem}\label{rem:cuspThydro}
Note that in the case of Hydrogen (\(N=1\)), with
\(\phi(x)=e^{\frac{Z}{2}|x|}\psi\),
\begin{align}
  \label{eq:cuspThydro}
  \widetilde t{}\,'(0)+Z\widetilde t(0)=  
  \frac{Z}{3}\Big(\big[\frac{Z^2}{4}+E\big]\widetilde\rho_H(0) 
  +4\pi|\nabla\phi(0)|^2\Big)\,,
\end{align}
with \(\nabla\phi\equiv0\) for the ground state.
\end{rem}
\subsection{Integrals and limits}\label{sec:int&limits}
In the proofs of Proposition~\ref{prop:cusps} and
Proposition~\ref{prop:t'} we shall restrict ourselves to proving the
existence of the  limits of the mentioned derivatives as
\(r\downarrow0\) (e.g., \(\lim_{r\downarrow0}v_1'(r,\omega)\)). The
existence of the limits of the difference quotients (here,
\(v_1'(0,\omega):=\lim_{r\downarrow0}\frac{v_1(r,\omega)-v_1(0)}{r}\)),
and their equality with the limits of the derivatives (i.e.,
\(\lim_{r\downarrow0}v_1'(r,\omega)\)), is a consequence of the
following lemma, which is an easy consequence of the mean value
theorem. 
\begin{lemma}\label{lem:meanValue}
Let \(f:[a,b]\to\R\) satisfy \(f\in C^0\big([a,b]\big)\), \(f'\in
C^0\big((a,b)\big)\), and that
\(\lim_{\epsilon\downarrow0}f'(a+\epsilon)\) exists. 

Then \(\lim_{\epsilon\downarrow0}\frac{f(a+\epsilon)-f(a)}{\epsilon}\)
exists and equals   \(\lim_{\epsilon\downarrow0}f'(a+\epsilon)\). 
\end{lemma}
Verifying the two first assumptions in Lemma~\ref{lem:meanValue} in
the case of the functions in Proposition~\ref{prop:cusps} and
Proposition~\ref{prop:t'} follow exactly the same ideas as proving the
existence of the limits of the mentioned derivatives as
\(r\downarrow0\) (which follows below), and so we will omit the
details. 

When proving below the existence of the  limits of the mentioned
derivatives as \(r\downarrow0\), we shall need to interchange first
the differentiation \(\frac{d}{dr}\), then the limit
\(\lim_{r\downarrow0}\), with the integration \(\int_{\mathbb
  R^{3N-3}}\cdots\,d\hat{\bf x}_1\) (or, for the spherical averages,
with \(\int_{\mathbb{S}^2}\int_{\mathbb R^{3N-3}}\cdots\,d\hat{\bf
  x}_1d\omega\)). We shall use Lebesgue's Theorem of Dominated
Convergence in both cases; in the case of \(\frac{d}{dr}\) in the form
of Proposition~\ref{prop:genLeb} below, which is a standard result in
integration theory (see e.\,g.\,\cite{Bony}). The dominant (in
Proposition~\ref{prop:genLeb}, the function \(g\)) will be the same in
both cases. 
\begin{prop}\label{prop:genLeb}
Let \(I\subset\mathbb R\) be an interval, \(A\) any subset of
\(\mathbb R^m\), and \(f\) a function defined on \(A\times I\),
satisfying the three following hypothesis: 
\begin{itemize}
  \item[i)]  For all \(\lambda\in I\), the function \(x\mapsto
             f(x,\lambda)\) is integrable on  \(A\).
  \item[ii)] The partial derivative \(\partial f/\partial
             \lambda(x,\lambda)\) exists at all points in \(A\times I\).
   \item[iii)] There exists a non-negative function \(g\), integrable on
               \(A\), such that \(|\partial f/\partial
               \lambda(x,\lambda)|\leq g(x)\) for all \((x,\lambda)\in
               A\times I\). 
  \end{itemize}
  
Then the function \(F\) defined by 
\begin{align*}
  F(\lambda)=\int_A f(x,\lambda)\,dx
\end{align*}
is differentiable in \(I\), and 
\begin{align*}
   F'(\lambda)=\int_A\frac{\partial f}{\partial \lambda}(x,\lambda)\,dx.
\end{align*}
\end{prop}
\begin{rem}\label{rem:exepSet}
Note that one can remove a set \(B\) of measure zero from the domain
of integration \(A\), without changing the two integrals above; it is
therefore enough to check the three hypothesis on this new domain,
\(A'=A\setminus B\). Note that the set \(B\) of measure zero must be
{\rm independent} of \(\lambda\).  
\end{rem}

Note that in \eqref{def:t}--\eqref{def:rho1} we can, for
\(x=r\omega\neq0\), restrict integration to the set
\begin{align}
  \label{eq:goodSet}
  \mathcal{S}_1(\omega)=\Big\{\hat{\bf x}_1\in\R^{3N-3}\,\Big|\,
  &\,x_k\neq0\,,\,
  \frac{x_k}{|x_k|}\neq\omega\,,\,
  x_\ell\neq x_k\,,
  \\&\,
  \text{ for }
  \,k,\ell\in \{2,\ldots,N\}
  \text{ with }\, k\neq\ell\,
  \Big\}\,,
  \nonumber
\end{align}
since \(\mathbb{R}^{3N-3}\setminus\mathcal{S}_1(\omega)\) has measure
zero. The set \(\mathcal{S}_1(\omega)\) will be our  \(A'\). From the
definition of \(\mathcal{S}_1(\omega)\) it follows that for any
\(r>0\), \(\omega\in \mathbb{S}^2\), and \(\hat{\bf
  x}_1\in\mathcal{S}_1(\omega)\), there exists an \(\epsilon>0\) and a
neighbourhood \(U\subset\mathcal{S}_1(\omega)\subset\R^{3N-3}\) of
\(\hat{\bf x}_1\) such that \(V\) in \eqref{eq:V} is \(C^\infty\) on
\(B_3(r\omega,\epsilon)\times U\subset \R^{3N}\). It follows from
elliptic regularity \cite{GandT} that \(\psi\in
C^{\infty}(B_3(r\omega,\epsilon)\times U)\). In particular, if
\(G:\R^{3N}\to\R\) is any of the integrands in
\eqref{def:t}--\eqref{def:rho1}, then, for all
\(\omega\in\mathbb{S}^2\), the partial derivative \(\partial
G_{\omega}/\partial r(r,\hat{\bf x}_1)\) of the function \((r,\hat{\bf
  x}_1)\mapsto G(r\omega,\hat{\bf x}_1)\equiv G_{\omega}(r,\hat{\bf
  x}_1)\) exists, and satisfies 
\begin{align}
  \label{eq:radial-der}
  \frac{\partial G_{\omega}}{\partial r}(r,\hat{\bf x}_1)=\omega\cdot\big[(\nabla_1
  G)(r\omega,\hat{\bf x}_1)\big] \ \text{ for all } r>0\,, \hat{\bf
  x}_1\in\mathcal{S}_1(\omega).
\end{align}
This assures that the hypothesis ii) in Proposition~\ref{prop:genLeb}
will be satisfied in all cases below where we apply  the proposition
(the hypothesis i) is trivially satisfied). 

We illustrate how to apply this. Let
\begin{align*}
  g_{\omega}(r)=g(r,\omega)=\int_{\mathcal{S}_1(\omega)}G(r\omega,\hat{\mathbf 
  x}_1)\, d\hat{\bf x}_1\ \ 
  (\ =\int_{\mathbb R^{3N-3}}
  G(r\omega,\hat{\mathbf 
  x}_1)\, d\hat{\bf x}_1 )\,.
\end{align*}
Interchanging integration and differentiation as discussed above, we
have, for \(\omega\in\mathbb{S}^2\) fixed and \(r>0\), that  
\begin{align}
  \label{eq:derDefault}
  g_{\omega}{}\!'(r)
  =\int_{\mathcal{S}_1(\omega)}\omega\cdot\big[(\nabla_1 G)(r\omega,\hat{\mathbf 
  x}_1)\big]\, d\hat{\bf x}_1.
\end{align}

To justify this, and to prove the existence of \(\lim_{r\downarrow0}
g_{\omega}{}\!'(r)\) we will need to prove two things. 

First, we need to find a dominant to the integrand in
\eqref{eq:derDefault}, that is, a function \(\Phi_{\omega}\in
L^{1}(\mathbb R^{3N-3})\) such that, for some \(R_{0}>0\), 

\begin{align}
  \label{eq:dominant}
  \Big|\omega\cdot\big(\nabla_1 G\big)
  (r\omega,\hat{\bf x}_1)\big)\Big|  
  \leq \Phi_{\omega}(\hat{\bf x}_1)
  \text{ for all } r\in(0,R_{0}), 
  \hat{\bf x}_1\in\mathcal{S}_1(\omega).
\end{align}
This will, by Proposition~\ref{prop:genLeb}, also justify the above
interchanging of \(\frac{d}{dr}\) and the integral
(\eqref{eq:dominant} ensures that the hypothesis iii) is
satisfied). We note that, with one exception, whenever we apply this,
in fact \(\Phi\equiv\Phi_{\omega}\) will be {\it independent} of
\(\omega\in\mathbb{S}^2\), and therefore \(\Phi\in L^{1}(\mathbb
R^{3N-3}\times\mathbb{S}^2)\).

Secondly, we need to prove, for all \(\omega\in\mathbb{S}^2\) and
\(\hat{\bf x}_1\in\mathcal{S}_1(\omega)\) fixed, the existence of 
\begin{align}
  \label{eq:lim_rTo0}
  \lim_{r\downarrow0}
  \Big[\omega\cdot\big(\nabla_1 G\big)
  (r\omega, \hat{\bf x}_1)\Big]. 
\end{align}
This, by Lebesgue's Theorem of Dominated Convergence, will prove the
existence of \(\lim_{r\downarrow0} g_{\omega}{}\!'(r)\). 

To study
\begin{align}
  \label{eq:tildeG}
  \widetilde g(r)=\int_{\mathbb{S}^2}g(r,\omega)\,d\omega
  =\int_{\mathbb{S}^2}\int_{\R^{3N-3}}G(r\omega,\hat{\mathbf x}_1)\,
  d\hat{\bf x}_1\,d\omega 
\end{align}
note that, by the above, the set
\begin{align*}
  \Big(\R^{3N-3}\times\mathbb{S}^2\Big)\setminus
  \Big(\bigcup_{\omega\in\mathbb{S}^2}\big[\mathcal{S}_1(\omega)
  \times\{\omega\}]\Big)
\end{align*}
has measure zero, and that, also by the above, the partial
derivative\par\noindent \(\partial \widetilde G/\partial r(r,(\hat{\bf
  x_1},\omega))\) of the function
\begin{align*}
  (r,(\hat{\bf x}_1,\omega))\mapsto
  \widetilde G(r,(\hat{\bf x}_1,\omega))\equiv G(r\omega,\hat{\bf
  x}_1)
\end{align*}
exists for all \(r>0\) and \((\hat{\bf x}_1,\omega)\in\cup_{\omega
  \in\mathbb{S}^2}\big[\mathcal{S}_1(\omega) \times\{\omega\}\big]\).
As noted above, the dominants \(\Phi\) we will exhibit below when
studying \(g(r,\omega)\) will (except in one case) be independent of
\(\omega\in\mathbb{S}^2\), and so can also be used to apply both
Lebegue's Theorem on Dominated Convergence,
Proposition~\ref{prop:genLeb}, and Fubini's Theorem on the integral in
\eqref{eq:tildeG}. This implies that
\begin{align}
  \label{eq:limitTildeG}
  \lim_{r\downarrow0}\widetilde{g}{}\,'(r)
  = \int_{\R^{3N-3}}\int_{\mathbb{S}^2}
  \Big\{\lim_{r\downarrow0}
  \omega\cdot\big[(\nabla_1G)(r\omega,\hat{\mathbf 
  x}_1)\big]\Big\}\,d\omega\,d\hat{\bf x}_1\,,
\end{align}
as soon as we have proved the pointwise convergence of the integrand
for all \(\omega\in\mathbb{S}^2\) and \(\hat{\bf
  x}_1\in\mathcal{S}_1(\omega)\), and provided the mentioned
dominant. (In the last, exceptional case, we will provide a dominant
\(\Phi\in L^1(\R^{3N-3}\times\mathbb{S}^2)\)).

In the sequel, we shall use all of this without further mentioning,
apart from proving the existence of a dominant, and the existence of
the pointwise limits of the integrands on
\(\mathcal{S}_1(\omega)\). Also, for notational convenience, we shall
allow ourselves to write all integrals over \(\R^{3N-3}\) instead of
over \(\mathcal{S}_1(\omega)\). 

\subsection{Additional partial regularity}
For the existence of pointwise limits the following lemma will be essential; it gives
detailed information about the structure of the eigenfunction \(\psi\)
in the vicinity of the two-particle singularity \(x=0\).
The lemma is reminiscent of more detailed results obtained earlier,
see \cite[Proposition 2]{CMP1}, \cite[Lemma 2.2]{taxco}, and
\cite[Lemma 3.1]{Ark}.

We need to recall the definition of H\"older-continuity.
\begin{defn}
  \label{def:Holder}
  Let \(\Omega\) be a domain in \(\R^{n}\), \(k\in\N\), and
  \(\alpha\in(0,1]\). We say that a function \(u\) belongs to
  \(C^{k,\alpha}(\Omega)\) whenever
  \(u\in C^{k}(\Omega)\), and for all \(\beta\in\N^{n}\) with
  \(|\beta|=k\), and all open balls $B_{n}(x_{0},r)$ with
  $\overline{B_{n}(x_{0},r)}\subset\Omega $, we have
  \begin{align*}
  \sup_{x,y\in B_{n}(x_{0},r),\,x\neq y}
  \!\!\!\!\!\!\!\!\!
  \frac{|D^{\beta}u(x)-D^{\beta}u(y)|}{|x-y|^{\alpha}}
  \leq C(x_{0},r).
\end{align*}
When \(k=0\) and \(\alpha\in(0,1)\) we also write
\(C^{\alpha}(\Omega)\equiv C^{0,\alpha}(\Omega)\).
\end{defn}
\begin{lemma}\label{lem:phi}
Let \(\omega\in\mathbb{S}^2\) and \(\hat{\bf
  x}^0_1\in\mathcal{S}_1(\omega)\). Then there exists an open
neighbourhood \(U\subset\mathcal{S}_1(\omega)\subset\R^{3N-3}\) of
\(\hat{\bf x}^0_1\) and \(\epsilon>0\) such that 
\begin{align}
  \label{eq:regPhi}
  \psi(x,{\hat{\bf x}_1})&=e^{-\frac{Z}{2}|x|}\varphi_1(x,{\hat{\bf
      x}_1})\ \text{ with } \\ 
  \label{eq:diffSing}
  \partial_{\hat{\bf x}_1}^\beta\varphi_1&\in
  C^{1,\alpha}\big(B_3(0,\epsilon)\times U\big)\ \text{ for all }
  \alpha\in(0,1), \beta\in\N^{3N-3}.
\end{align}
\end{lemma}
\begin{proof}
By the definition \eqref{eq:goodSet} of \(\mathcal{S}_1(\omega)\)
there exists a neighbourhood
\(U\subset\mathcal{S}_1(\omega)\subset\R^{3N-3}\) of \(\hat{\bf
  x}^0_1\in\mathcal{S}_1(\omega)\), and \(\epsilon>0\) such that 
\begin{align}\label{eq:niceSet}
  &x_j \neq x_k \qquad\qquad\, \text{ for } j,k \in
  \{2,\ldots,N\} \,\text{ with } j\neq k\,,\\
  \nonumber
  &x_j \neq 0\,,\, x_j \neq x \quad \text{ for } j \in \{2, \ldots,
  N\}\\\nonumber &\text{for all } (x,x_2,\ldots,x_N)\in B_3(0,\epsilon)\times
  U\subset\R^{3N}.\nonumber 
\end{align}
Make the {\it Ansatz} \(\psi=e^{-\frac{Z}{2}|x|}\varphi_1\). Using
\eqref{Hpsi}, \eqref{Hbis}, and that \(\Delta_x(|x|)=2|x|^{-1}\), we get
that \(\varphi_1\) satisfies the equation
\begin{align}
  \label{eq:phi1}
  \Delta\varphi_1-Z\frac{x}{|x|}\cdot\nabla_1\varphi_1+(\tfrac{Z^2}{4}+E-V_1)\varphi_1=0 
\end{align}
with
\begin{align}
  \label{def:V1}
  V_1(x,\hat{\bf x}_1)=\sum_{j=2}^{N}-\frac{Z}{|x_j|} +
  \sum_{k=2}^N\frac{1}{|x-x_k|}+
  \sum_{2\leq j<k\leq
    N}\frac{1}{|x_j-x_k|}\, ,
\end{align}
where, due to \eqref{eq:niceSet}, 
\begin{align}
  V_1\in C^{\infty}(B_3(0,\epsilon)\times U).
\end{align}
Since the coefficients in \eqref{eq:phi1} are in
\(L^{\infty}(B_3(0,\epsilon)\times U))\), this implies, by standard elliptic
regularity \cite[Theorem 8.36]{GandT}, 
that 
\begin{align}
  \label{eq:firstPhi1}
  \varphi_1\in C^{1,\alpha}(B_3(0,\epsilon)\times U) \text{ for all }
  \alpha\in(0,1).
\end{align}
Let \(\eta\in \mathbb{N}^{3N-3}\), \(|\eta|=1\); differentiating
\eqref{eq:phi1} we get
\begin{align}
  \label{eq:secPhi1}
  \Delta(\partial^{\eta}_{\hat{\bf x}_1}
  \varphi_1)-Z\frac{x}{|x|}&\cdot\nabla_1(\partial^{\eta}_{\hat{\bf
      x}_1}\varphi_1)
  \\
  &=-\big[(\tfrac{Z^2}{4}+E-V_1)(\partial^{\eta}_{\hat{\bf
      x}_1}\varphi_1)-(\partial^{\eta}_{\hat{\bf x}_1}
  V_1)\varphi_1\big]\,. 
  \nonumber
\end{align}
By \eqref{def:V1} and \eqref{eq:firstPhi1}, the right side in
\eqref{eq:secPhi1} belongs to \(C^{\alpha}(B_3(0,\epsilon)\times U)\)
for all \(\alpha\in(0,1)\)
and so, by elliptic regularity, \(\partial^{\eta}_{\hat{\bf x}_1} 
  \varphi_1\in C^{1,\alpha}(B_3(0,\epsilon)\times U)\) for all
  \(\alpha\in(0,1)\). An easy induction 
  argument finally gives that  
\begin{align*}
  \partial^{\beta}_{\hat{\bf x}_1}
  \varphi_1\in C^{1,\alpha}(B_3(0,\epsilon)\times U) \text{ for all
  }\alpha\in(0,1), \beta\in\mathbb{N}^{3N-3}. 
\end{align*}
\end{proof}
\subsection{Proof of Proposition~\ref{prop:cusps}}
We will treat the two kinds of terms in \eqref{def:w} separately. 

For the first one, we make a change of variables. Assume without loss
of generality that \(k=2\), and let \(y=x_2-r\omega\), then 
\begin{align}
  \label{eq:var-change}
  \int_{\R^{3N-3}}&\frac{1}{|r\omega-x_2|}\,|\psi(r\omega,\hat{\bf
    x}_1)|^2\,d\hat{\bf x}_1
  \\&=\int_{\R^{3N-3}}\frac{1}{|y|}\,|\psi(r\omega,y+r\omega,x_3,\ldots,x_N)|^2
  \,dy\,dx_3\ldots dx_N\,. 
  \nonumber
\end{align}
For \(\omega\in\mathbb{S}^2\) fixed, and \(r>0\), interchanging
integration and differentiation as discussed above, we get, using
\eqref{eq:var-change}, that
\begin{align}
  \label{eq:der-var-change}
  \nonumber
  &\frac{d}{dr}\Big[\int_{\R^{3N-3}}\frac{1}{|r\omega-x_2|}\,|\psi(r\omega,\hat{\bf
  x}_1)|^2\,d\hat{\bf x}_1\Big]
  \\&=\int_{\R^{3N-3}}\Big[\frac{2}{|y|}\,
  \psi\big\{\omega\cdot
  (\nabla_1\psi+\nabla_2\psi)\big\}\Big](r\omega,y+r\omega,x_3,\ldots,x_N)\big)
  \\&\qquad\qquad\qquad\qquad\qquad\qquad\qquad\qquad\qquad\qquad\qquad
  \,dy\,dx_3\ldots dx_N. \nonumber
\end{align}
Note that, using \eqref{eq:exp-dec}, \eqref{eq:dec_grad_psi}, and
equivalence of norms in \(\R^{3N}\), there exists positive constants
\(C, c_1, c_2\) such that 
\begin{align}
  \label{eq:dominantChangeVar}
  \nonumber
  \Big|\frac{2}{|y|}\,
  \psi\big\{\omega\cdot
  (\nabla_1\psi+\nabla_2\psi)\big\}(r\omega,y&+r\omega,x_3,\ldots,x_N)\Big|
  \\&
  \le C\,e^{c_1r}\frac{1}{|y|}\,e^{-c_2|(y,x_3,\ldots,x_N)|},
\end{align}
which provides a dominant, independent of \(\omega\in\mathbb{S}^2\),
in the sense of \eqref{eq:dominant}, uniformly for \(r\in(0,R_0)\) for
any \(R_0>0\).

Writing \(\psi=e^{-\frac{Z}{2}|x|}\varphi_1\), the integrand in
\eqref{eq:der-var-change} equals 
\begin{align*}
  \frac{1}{|y|}
  \big[-Z|\psi|^2+2\varphi_1\,e^{-Zr}(\omega\cdot\{\nabla_1\varphi_1
  +\nabla_2\varphi_1\})\big]
  (r\omega,y+r\omega,x_3,\ldots,x_N),
\end{align*}
which has a limit as \(r\downarrow0\) (for \(\omega\in\mathbb{S}^2\)
and \((y,x_3,\ldots,x_N)\) fixed) since \(\varphi_1\in C^{1,\alpha}\)
by Lemma~\ref{lem:phi}. This proves the existence of the limit of the
integrand in \eqref{eq:der-var-change}, and therefore of the limit as
\(r\downarrow0\) of the derivative with respect to \(r\) of the first
term in \eqref{def:w}. Note that since \(\int_{\mathbb
  S^2}\omega\,d\omega=0\), the terms proportional to \(\omega\) vanish
by integration over \((y,x_3,\ldots,x_N,\omega)\). The limit of the
derivative of the spherical average of this term is therefore (after
setting \(x_2=y\)) 
\begin{align}
  \label{eq:firstTermW1}
  -Z\int_{\mathbb{S}^2}\int_{\R^{3N-3}}\frac{1}{|x_2|}\,|\psi(0,\hat{\bf
    x}_1)|^2\,d\hat{\bf x}_1\,d\omega\,.
\end{align}

For the second term  in \eqref{def:w}, assume without loss that
\(k=2\), \(\ell=3\), and write \(\psi= e^{-\frac{Z}{2}|x|}\varphi_1\)
as before. Then, for \(\omega\in\mathbb{S}^2\) fixed, and \(r>0\),
interchanging integration and differentiation we get
\begin{align}
  \label{eq:secondW2}
  &\frac{d}{dr}\Big[\int_{\R^{3N-3}}\frac{1}{|x_{2}-x_{3}|}\,|\psi(x, \hat{\bf 
  x}_1)|^{2}\,d\hat{\bf x}_1\Big]
  \nonumber
  \\&=    \int_{\mathbb R^{3N-3}}\frac{2}{|x_2-x_3|}
  \,\psi(r\omega,\hat{\bf x}_1)
  \big[\omega\cdot
  \nabla_1\psi(r\omega,\hat{\bf x}_1)\big]
  \, d\hat{\bf x_1}\,.
  \\&
  =\int_{\mathbb R^{3N-3}}
  \frac{1}{|x_{2}-x_{3}|}
  \big[-Z|\psi|^2 + 2\varphi_1\,e^{-Zr} 
       (\omega\cdot\nabla_1\varphi_1)
  \big](r\omega,\hat{\bf x}_1)
  \, d\hat{\bf x}_1.
  \label{eq:secondW2BIS}
\end{align}
As above, \eqref{eq:exp-dec} and \eqref{eq:dec_grad_psi} provides us
with a dominant to the integrand in \eqref{eq:secondW2} in the sense
of \eqref{eq:dominant}.

Also as before, the integrand in \eqref{eq:secondW2BIS} has a limit as
\(r\downarrow0\), since \(\varphi_1\in C^{1,\alpha}\) by
Lemma~\ref{lem:phi}. This proves the existence of the limit as
\(r\downarrow0\) of the derivative of the second term in
\eqref{def:w}. Again, the term in \eqref{eq:secondW2BIS} proportional
to \(\omega\) vanish when integrating over \((\hat{\bf x}_1,\omega)\),
since \(\int_{\mathbb S^2}\omega\,d\omega=0\). The limit of the
derivative of the spherical average of this term is therefore
\begin{align}
  \label{eq:secondTermW1}
  -Z\int_{\mathbb{S}^2}\int_{\R^{3N-3}}\frac{1}{|x_2-x_3|}\,|\psi(0,\hat{\bf
    x}_1)|^2\,d\hat{\bf x}_1\,d\omega\,.
\end{align}

This proves the existence of \(\lim_{r\downarrow0}
w_1{}\!'(r,\omega)\) (for any \(\omega\in\mathbb{S}^2\) fixed), and of
\(\lim_{r\downarrow0}\widetilde w_1{}\!'(r)\). Furthermore, from
\eqref{eq:firstTermW1} and \eqref{eq:secondTermW1}, 
\begin{align*}
  \lim_{r\downarrow0}\widetilde w_1{}\!'(r)
  &={}-Z\int_{\mathbb S^2}\int_{\mathbb
    R^{3N-3}}\Big[\sum_{k=2}^{N}\frac{1}{|x_{k}|}\Big]\,|\psi(0,\hat 
  {\bf x}_1)|^2\, d\hat{\bf x}_1 d\omega\\&
  \quad{}-Z\int_{\mathbb S^2}\int_{\mathbb
      R^{3N-3}}
  \Big[\sum_{2\leq k<\ell\leq N}\frac{1}{|x_{k}-x_\ell|}
  \Big]\,|\psi(0,\hat
  {\bf x}_1)|^2\, d\hat{\bf x}_1 d\omega
  \\&={}-Z\widetilde w_1(0).
\end{align*}

The proof for \(v_1\) is similar, but simpler; we give it for
completeness. For \(r>0\) we get, by arguments as for \(w_1{}\!'\),
that 
\begin{align}
  \label{eq:v_1'}
  &v_1{}\!'(r,\omega)
  =\sum_{k=2}^{N}\int_{\mathbb
  R^{3N-3}}
  \frac{2Z}{|x_{k}|}\,\psi(r\omega,\hat{\bf
  x}_1)\big[\omega\cdot\nabla_1\psi(r\omega,\hat{\bf x}_1)\big]
  \, d\hat{\bf x}_1 
  \\
  \label{eq:v_1'BIS}
  &=\sum_{k=2}^{N}\int_{\mathbb R^{3N-3}} 
  \frac{Z}{|x_{k}|}\big[ -Z|\psi|^2+2\varphi_1\,e^{-Zr}
  (\omega\cdot\nabla_1 \varphi_1)
  \big](r\omega,\hat{\mathbf x}_1)\, d\hat{\bf x}_1.
\end{align}
One provides a dominant to the integrand in \eqref{eq:v_1'} in a
similar way as for \(w_1\). We omit the details. Again, existence of
the limit as \(r\downarrow0\) of the integrand in  \eqref{eq:v_1'BIS}
is ensured by Lemma~\ref{lem:phi}. 

The last term in \eqref{eq:v_1'BIS} again vanishes when taking the
limit \(r\downarrow0\) and then integrating over \((\hat{\bf
  x}_1,\omega)\), since \(\int_{\mathbb S^2}\omega\,d\omega=0\), and
so 
\begin{align*}
  \lim_{r\downarrow0}\widetilde v_1{}\!'(r,\omega)  
  &={}-Z\sum_{k=2}^{N}\int_{\mathbb S^2}\int_{\mathbb
  R^{3N-3}}\frac{Z}{|x_{k}|}\,|\psi(0,\hat
  {\bf x}_1)|^2\, d\hat{\bf x}_1 d\omega
  \\&={}-Z\widetilde v_1(0).
\end{align*}
This finishes the proof of Proposition~\ref{prop:cusps}.
\qed
\subsection{Proof of Proposition~\ref{prop:t'}}
We proceed as in the proof of
Proposition~\ref{prop:cusps}. Interchanging integration and
differentiation as discussed in Section~\ref{sec:int&limits} we have,
for \(\omega\in\mathbb{S}^2\) fixed and \(r>0\), that
\begin{align}
  \label{eq:der1}
  t_1{}\!'(r,\omega)
  =\int_{\mathbb
    R^{3N-3}}\omega\cdot\big[\nabla_1\big(|\nabla\psi|^2\big)(r\omega,\hat{\mathbf
    x}_1)\big]\, d\hat{\bf x}_1.
\end{align}

Again, to justify this and to prove the existence of
\(\lim_{r\downarrow0} t_1{}\!'(r,\omega)\) we need to prove two
things: The existence of the pointwise limit as \(r\downarrow0\) of
the integrand above, for \(\omega\in\mathbb{S}^2\) and \(\hat{\bf
  x}_1\in\mathcal{S}_1(\omega)\) fixed, and the existence of a
dominant in the sense of \eqref{eq:dominant}. 

\subsection*{\bf Pointwise limits.}
We start with the pointwise limit. We allow ourselves to compute the
integrals of the found limits, presuming the dominant found. We shall
provide the necessary dominants afterwards. 

Recall from (the proof of) Lem\-ma~\ref{lem:phi} that
\(\varphi_1=e^{\frac{Z}{2}|x|}\psi\) satisfies the equation 
\begin{align}
  \label{eq:phi1BIS}
  \Delta\varphi_1-Z\frac{x}{|x|}\cdot\nabla_1\varphi_1
  +(\tfrac{Z^2}{4}+E-V_1)\varphi_1=0  
\end{align}
where
\begin{align*}
  V_1\in C^{\infty}(B_3(0,\epsilon)\times U)
\end{align*}
for some \(\epsilon>0\) and
\(U\subset\mathcal{S}_1(\omega)\subset\R^{3N-3}\) some neighbourhood
of \(\hat{\bf x}_1\). 

Using Lemma~\ref{lem:phi}, we get that \(\Delta_{\hat{\bf
    x}_1}\varphi_1 \in C^{1,\alpha}(B_3(0,\epsilon)\times U)\) for all
\(\alpha\in(0,1)\). From this and \eqref{eq:phi1BIS} follows that for
any \(\hat{\bf x}_1\in\mathcal{S}_1(\omega)\) fixed, and some
\(\epsilon>0\), the function \(x\mapsto \varphi_1(x,\hat{\bf x}_1)\)
satisfies the equation
\begin{align}
  \label{eq:newPhi1}
  \Delta_x\varphi_1-Z\frac{x}{|x|}\cdot\nabla_x\varphi_1
  &=-\Delta_{\hat{\bf x}_1}\varphi_1-(\tfrac{Z^2}{4}+E-V_1)\varphi_1
  \\&
  \equiv G_{\hat{\bf x}_1}\ , \ G_{\hat{\bf x}_1}\in
  C^{1,\alpha}(B_3(0,\epsilon))\,,\,\alpha\in(0,1)\,.
\end{align}
Furthermore, from Lemma~\ref{lem:phi} (with \(\beta=0\)) it follows
that \(\nabla_1\varphi_1(0,\hat{\bf x}_1)\) is well-defined. Note that
for \(c\in \R^3\) the function
\begin{align}
  v(x)=\frac14|x|^2(c\cdot\omega)
  =\frac14|x|(x\cdot c) 
\end{align}
solves \(\Delta_xv=c\cdot\omega\). Therefore, with
\(c=Z\nabla_1\varphi_1(0,\hat{\bf x}_1)\), the function
\(u=u_{\hat{\bf x}_1}=\varphi_1(\,\cdot\,,\hat{\bf x}_1)-v\) satisfies
the equation (in \(x\), with \(\hat{\bf x}_1\in\mathcal{S}_1(\omega)\)
fixed)
\begin{align}
  \label{eq:u}
  \nonumber
  \Delta_xu(x)&=
  Z\frac{x}{|x|}\cdot\big(\nabla_1\varphi_1(x,\hat{\bf x}_1)
  -\nabla_1\varphi_1(0,\hat{\bf x}_1)\big)
  \\&\ 
  -\Delta_{\hat{\bf x}_1}\varphi_1(x,\hat{\bf x}_1)-(\tfrac{Z^2}{4}+E-V_1)\varphi_1(x,\hat{\bf x}_1)
  \equiv g_{\hat{\bf x}_1}(x).
\end{align}
By the above, and Lemma~\ref{usefullemma} in Appendix~\ref{app:lem}
(using that \(\nabla_1\varphi_1\) is \(C^{\alpha}\)), \(g_{\hat{\bf
    x}_1}\in C^{\alpha}(\R^{3})\) for all \(\alpha\in(0,1)\), and so,
by standard elliptic regularity, \(u\in C^{2,\alpha}(\R^{3})\) for all
\(\alpha\in(0,1)\). To recapitulate, for any \(\hat{\bf
  x}_1\in\mathcal{S}_1(\omega)\), the function
\(x\mapsto\varphi_1(x,\hat{\bf x}_1)\) satisfies 
\begin{align}
  \label{eq:reprPhi}
  \varphi_1(x,\hat{\bf x}_1)=u_{\hat{\bf
  x}_1}(x)+\frac14|x|^2(c\cdot\frac{x}{|x|})\,, \ u_{\hat{\bf x}_1}\in
  C^{2,\alpha}(\R^3)\,,\,\alpha\in(0,1)\,.
\end{align}
Note that
\begin{align}
  \label{eq:valueC}
  c=Z\nabla_1\varphi_1(0,\hat{\bf x}_1)
  =Z\nabla_xu_{\hat{\bf x}_1}(0).
\end{align}

We now apply the above to prove the existence of the pointwise limit
of the integrand in \eqref{eq:der1} for fixed
\(\omega\in\mathbb{S}^2\) and \(\hat{\bf
  x}_1\in\mathcal{S}_1(\omega)\).

First, since \(\psi=e^{-\frac{Z}{2}|x|}\varphi_1\), we have, for
\(j=2,\ldots,N\), 
\begin{align*}
  \omega\cdot\big[\nabla_1\big(|\nabla_j\psi|^2\big)(r\omega, \hat{\bf
    x_1})\big]
  &=\omega\cdot\big[\nabla_1\big(e^{-Z|x|}|\nabla_j\varphi_1|^2\big)
  (r\omega,\hat{\bf x}_1)\big] 
  \\&=-Ze^{-Zr}|\nabla_j\varphi_1|^2(r\omega,\hat{\bf x}_1)
  \\&\ \ +e^{-Zr}\big\{\omega\cdot\big[\nabla_1
\big(|\nabla_j\varphi_1|^2\big)(r\omega,\hat{\bf x}_1)\big]\big\}. 
\end{align*}
Because of \eqref{eq:diffSing} in Lemma~\ref{lem:phi}, this has a
limit as \(r\downarrow0\) for fixed \(\omega\in\mathbb{S}^2\) (recall
that \(j=2,\ldots,N\)). The contribution to
\(\lim_{r\downarrow0}\widetilde t_1{}\!'(r)\) from this is 
\begin{align}
  \label{contr:t1a}
  {}-Z\sum_{j=2}^{N}\int_{\mathbb{S}^2}\int_{\R^{3N-3}}
  |\nabla_j\varphi_1(0,\hat{\bf x}_1)|^2
  \,d\hat{\bf x_1}\,d\omega\,,
\end{align}
since terms proportional with \(\omega\) vanish upon integration.

So we are left with considering
\(\omega\cdot\nabla_1\big(|\nabla_1\psi|^2\big)\) (see
\eqref{eq:der1}). To this end, use again
\(\psi=e^{-\frac{Z}{2}|x|}\varphi_1\) to get 
\begin{align}
 \label{eq:last-term1}
  \omega\cdot\nabla_x\big(|\nabla_1&\psi|^2\big)
  =\omega\cdot\nabla_x\big(\big|-\tfrac{Z}{2}\omega\psi
  +e^{-\frac{Z}{2}|x|}\nabla_1\varphi_1 \big|^2\big)
  \\&=\omega\cdot\nabla_x\big(\tfrac{Z^2}{4}\psi^2
  +e^{-Z|x|}|\nabla_1\varphi_1|^2-Z(\omega\cdot\nabla_1\varphi_1)
  e^{-Z|x|}\varphi_1\big).
  \nonumber
\end{align}
(We leave out the variables, \((r\omega,\hat{\bf x}_1)\)). We will
study each of the three terms in \eqref{eq:last-term1} separately.

For the first term in \eqref{eq:last-term1}, again using
\(\psi=e^{-\frac{Z}{2}|x|}\varphi_1\), we have
\begin{align}
  \label{eq:firstTerm}
  \omega\cdot\nabla_x\big(\tfrac{Z^2}{4}\psi^2\big)
  =-\tfrac{Z^3}{4}\psi^2+\tfrac{Z^2}{2}\psi
  e^{-\frac{Z}{2}|x|}(\omega\cdot\nabla_1\varphi_1), 
\end{align}
which has a limit as \(r\downarrow0\) for fixed
\(\omega\in\mathbb{S}^2\) and \(\hat{\bf
  x}_1\in\mathcal{S}_1(\omega)\), since \(\varphi_1\) is
\(C^{1,\alpha}\) (see \eqref{eq:firstPhi1}). The contribution to
\(\lim_{r\downarrow0}\widetilde t_1{}\!'(r)\) from this is 
\begin{align}
  \label{contr:t1b}
  {}-\frac{Z^3}{4}\int_{\mathbb{S}^2}\int_{\R^{3N-3}}|\psi(0,\hat{\bf
    x}_1)|^2\,d\hat{\bf x_1}\,d\omega
  ={}-\frac{Z^3}{4}\widetilde\rho_1(0)\,.
\end{align}

As for the second term in \eqref{eq:last-term1} we have
\begin{align}
  \label{eq:secondBIS}
  \omega\cdot\nabla_x\big(e^{-Z|x|}|\nabla_1\varphi_1|^2\big)
  =-Ze^{-Z|x|}|\nabla_1\varphi_1|^2+e^{-Z|x|}(\omega\cdot\nabla_1|\nabla_1\varphi_1|^2), 
\end{align}
where the first term has a  limit as \(r\downarrow0\) for fixed
\(\omega\in\mathbb{S}^2\) and \(\hat{\bf
  x}_1\in\mathcal{S}_1(\omega)\), since \(\varphi_1\) is
\(C^{1,\alpha}\) (see \eqref{eq:firstPhi1}). This contributes
\begin{align}
  \label{contr:t1c}
  {}-Z\int_{\mathbb{S}^2}\int_{\R^{3N-3}}
  |\nabla_1\varphi_1(0,\hat{\bf x}_1)|^2
  \,d\hat{\bf x_1}\,d\omega\,
\end{align}
to \(\lim_{r\downarrow0}\widetilde t_1{}\!'(r)\). For the second term
(in \eqref{eq:secondBIS}), using that
\(\varphi_1=u+\frac{1}{4}r^2(c\cdot\omega)\) (see \eqref{eq:reprPhi}),
we get that 
\begin{align*}
  |\nabla_1\varphi_1|^2&=|\nabla_1u|^2
  +\frac12\big[(\omega\cdot\nabla_1u)(c\cdot
  x)+r(c\cdot\nabla_1u)\big]
  \\&\ \ +\frac{3}{16}(c\cdot x)^2
  +\frac{1}{16}r^2(c\cdot c),
\end{align*}
and so
\begin{align*}
  \omega\cdot\nabla_1|\nabla_1\varphi_1|^2
  &=2\langle\omega,(D^2u)\nabla_1u\rangle
  +\frac38r(c\cdot\omega)^2
  +\frac{1}{8}r(c\cdot c)
  \\&\ \ +\frac12\big[\langle\omega,(D^2u)\omega\rangle(c\cdot
  x)+(\omega\cdot\nabla_1u)(c\cdot\omega)
  \\&\qquad\qquad\qquad\qquad
  +(c\cdot\nabla_1u)
  +r\langle\omega,(D^2u)c\rangle\big]\,.
\end{align*}
Here, \(D^2u\) is the Hessian matrix of \(u=u_{\hat{\bf x}_1}\) with
respect to \(x\), and \(\langle\,\cdot,\cdot\rangle\) is the scalar
product in \(\R^3\). Since \(\varphi_1\in C^{1,\alpha}\), and \(u\in
C^{2,\alpha}\), all terms have a  limit as \(r\downarrow0\) for fixed
\(\omega\in\mathbb{S}^2\) and \(\hat{\bf
  x}_1\in\mathcal{S}_1(\omega)\). We get 
\begin{align}
  \label{eq:second2}\nonumber
  \lim_{r\downarrow0}\big[(e^{-Z|x|}\omega\cdot\nabla_1
  &|\nabla_1\varphi_1|^2)(r\omega,\hat{\bf x}_1)\big] 
  =2\omega\cdot\big[(D^2u_{\hat{\bf x}_1})(0)\nabla_xu_{\hat{\bf
      x}_1}(0)\big]
  \\&
  +\frac12\big[(c\cdot\nabla_xu_{\hat{\bf x}_1}(0))
  +(\omega\cdot\nabla_xu_{\hat{\bf x}_1}(0))(c\cdot\omega)\big].
\end{align}
When integrating, this contributes
\begin{align}
  \label{contr:t1d}
   \frac{2Z}{3}\int_{\mathbb{S}^2}\int_{\R^{3N-3}}
   |\nabla_1\varphi_1(0,\hat{\bf x}_1)|^2
   \,d\hat{\bf x_1}\,d\omega\,
\end{align}
to  \(\lim_{r\downarrow0}\widetilde t_1{}\!'(r)\); to see this, use
\eqref{eq:valueC} and Lemma~\ref{eq:dotprod} in Appendix~\ref{app:A},
and that \(\int_{\mathbb{S}^2}\omega\,d\omega=0\). 

For the third and last term in \eqref{eq:last-term1}, using that
\(\varphi_1=u+\frac{1}{4}r^2(c\cdot\omega)\) (see \eqref{eq:reprPhi})
and \(\omega\cdot\nabla_x=\partial_{r}\), we get that
\begin{align*}
  \omega\cdot\nabla_x&\big(-Z(\omega\cdot\nabla_1\varphi_1)e^{-Z|x|}\varphi_1\big) 
  =Z^2(\omega\cdot\nabla_1\varphi_1)e^{-Z|x|}\varphi_1
  \\&\ -Ze^{-Z|x|}\varphi_1\big[\,\langle\omega,(D^2 u) 
  \omega\rangle+\tfrac12(c\cdot\omega)\,\big]
  -Z(\omega\cdot\nabla_x\varphi_1)^2e^{-Z|x|}\,.
\end{align*}
Again, since \(\varphi_1\in C^{1,\alpha}\), and \(u\in
C^{2,\alpha}\), all terms have a  limit as \(r\downarrow0\) for fixed
\(\omega\in\mathbb{S}^2\) and \(\hat{\bf
  x}_1\in\mathcal{S}_1(\omega)\).

The contribution from this to  \(\lim_{r\downarrow0}\widetilde
t_1{}\!'(r)\) is, using Lemma~\ref{eq:dotprod}, and
\(\int_{\mathbb{S}^2}\omega\,d\omega=0\), 
\begin{align}
  \label{contr:t1e}
  &{}-\frac{Z}{3}\int_{\mathbb{S}^2}\int_{\R^{3N-3}}
  |\nabla_1\varphi_1(0,\hat{\bf x}_1)|^2
  \,d\hat{\bf x_1}\,d\omega\,
  \\&\qquad\qquad -\frac{Z}{3}\int_{\mathbb{S}^2}\int_{\R^{3N-3}}
  \Delta_xu_{\hat{\bf x}_1}(0)
  \varphi_1(0,\hat{\bf x}_1)
  \,d\hat{\bf x_1}\,d\omega\,.
  \nonumber
\end{align}
Here we used that \(\Tr(D^2f)=\Delta f\). 

This proves the existence of the pointwise limit of the integrand in
\eqref{eq:der1} for fixed \(\omega\in\mathbb{S}^2\) and \(\hat{\bf
  x}_1\in\mathcal{S}_1(\omega)\). 
Also, from  \eqref{contr:t1a}, \eqref{contr:t1b},  \eqref{contr:t1c}
\eqref{contr:t1d}, and \eqref{contr:t1e},
\begin{align}
  \label{eq:limitT'}
  \nonumber
  \lim_{r\downarrow0}\widetilde t_1{}\!'(r)
  ={}&-\Big[Z\int_{\mathbb{S}^2}\int_{\R^{3N-3}}
  |\nabla\varphi_1(0,\hat{\bf x}_1)|^2
  \,d\hat{\bf x}_1\,d\omega\,
  +\frac{Z^3}{4}\widetilde\rho_1(0)\Big]
  \\\nonumber
  & +\frac{Z}{3}\int_{\mathbb{S}^2}\int_{\R^{3N-3}}
  |\nabla_1\varphi_1(0,\hat{\bf x}_1)|^2
  \,d\hat{\bf x}_1\,d\omega\,
  \\&-\frac{Z}{3}\int_{\mathbb{S}^2}\int_{\R^{3N-3}}
  \Delta_xu_{\hat{\bf x}_1}(0)
  \varphi_1(0,\hat{\bf x}_1)
  \,d\hat{\bf x}_1\,d\omega\,.
\end{align}
Note that, due to \eqref{eq:u} and
\(\psi=e^{-\frac{Z}{2}|x|}\varphi_1\),
\begin{align*}
  \Delta_xu_{\hat{\bf x}_1}(0)={}-\Delta_{\hat{\bf
  x}_1}\psi(0,\hat{\bf x}_1)
  -\big(\tfrac{Z^2}{4}+E-V_{N-1,Z-1}(\hat{\bf
    x}_1)\big)\psi(0,\hat{\bf x}_1)\,,
\end{align*}
since \(V_1(0,\hat{\bf x}_1)=V_{N-1,Z-1}(\hat{\bf x}_1)\) (see
\eqref{eq:V}). This implies that
\begin{align}
  \label{eq:t'-a} 
  \nonumber
  \lim_{r\downarrow0}\,&\widetilde t_1{}\!'(r)
  ={}-\Big[Z\int_{\mathbb{S}^2}\int_{\R^{3N-3}}
  |\nabla\varphi_1(0,\hat{\bf x}_1)|^2
  \,d\hat{\bf x}_1\,d\omega\,
  +\frac{Z^3}{4}\widetilde\rho_1(0)\Big]
  \\
  &+\frac{Z}{3}\Big[\int_{\mathbb{S}^2}\int_{\R^{3N-3}}
  |\nabla_1\varphi_1(0,\hat{\bf x}_1)|^2
  \,d\hat{\bf x}_1\,d\omega
  +\frac{Z^2}{4}\widetilde\rho_1(0)
  \\&
  \!\!\!\!\!\!\!\!\!\!\!\!
  -\int_{\mathbb{S}^2}\int_{\R^{3N-3}}
  \Big[|\nabla_{\hat{\bf x}_1}\psi(0,\hat{\bf
  x}_1)|^2+\big(V_{N-1,Z-1}(\hat{\bf x}_1)-E\big)|\psi(0,\hat{\bf x}_1)|^2\Big]\,   
  \,d\hat{\bf x}_1\,d\omega\Big].
  \nonumber
\end{align}
Here we used that
\begin{align}
   {}-\int_{\R^{3N-3}}
   \psi(0,\hat{\bf x}_1)\big(\Delta_{\hat{\bf x}_1}\psi\big)(0,\hat{\bf x}_1)
   \,d\hat{\bf x}_1    
   =\int_{\R^{3N-3}}
   |\nabla_{\hat{\bf x}_1}\psi(0,\hat{\bf x}_1)|^2
   \,d\hat{\bf x}_1\,.
\end{align}

Before we provide a dominant in the sense of \eqref{eq:dominant} to
the integrand in \eqref{eq:der1} we now compute \(\widetilde
t_1(0)\). Note that for \(r>0\), 
\begin{align}
  \label{eq:t1}
  \widetilde t_1(r)&=
  \int_{\mathbb
    S^2}\int_{\R^{3N-3}}|\nabla_1\psi(r\omega,\hat{\bf
  x}_1)|^{2}\,d\hat{\bf x}_1\,d\omega
  \\&\qquad\qquad
  +\sum_{j=2}^{N}
  \int_{\mathbb
    S^2}\int_{\R^{3N-3}}|\nabla_j\psi(r\omega,\hat{\bf
  x}_1)|^{2}\,d\hat{\bf x}_1\,d\omega.
  \nonumber
\end{align}
Use \(\psi=e^{-\frac{Z}{2}|x|}\varphi_1\), then, for all
\(\omega\in\mathbb{S}^2\) and \(\hat{\bf
  x}_1\in\mathcal{S}_1(\omega)\) fixed, and \(j=2,\ldots,N\),
Lemma~\ref{lem:phi} gives that 
\begin{align}
   \label{eq:t1firstTerm}
   &|\nabla_{j}\psi(r\omega,\hat{\bf x}_1)|^{2}
   =e^{-Zr}|\nabla_{j}\varphi_1(r\omega,\hat{\bf x}_1)|^{2}
   \quad\overset{r\to0}{\longrightarrow}\quad
   |\nabla_{j}\varphi_1(0,\hat{\bf x}_1)|^{2}\,.
\end{align}
In particular, this proves the existence of 
\begin{align}
  \label{eq:partKin}
  |\nabla_{\hat{\bf x}_1}\psi(0,\hat{\bf x}_1)|^2
  =\sum_{j=2}^N|\nabla_j\psi(0,\hat{\bf x}_1)|^2
  =\sum_{j=2}^N|\nabla_j\varphi_1(0,\hat{\bf x}_1)|^2\,.
\end{align}
Similarly, 
\begin{align}
  \label{eq:secondTermT1}
   &|\nabla_{1}\psi(r\omega,\hat{\bf x}_1)|^{2}
   =\frac{Z^2}{4}|\psi(r\omega,\hat{\bf x}_1)|^2
   +e^{-Zr}|\nabla_1\varphi_1(r\omega,\hat{\bf x}_1)|^2
   \nonumber
   \\&\qquad\qquad-Ze^{-\frac{Z}{2}r}\psi(r\omega,\hat{\bf
   x}_1)\big(\omega\cdot\nabla_1\varphi_1(r\omega,\hat{\bf x}_1)\big)
   \\&\overset{r\to0}{\longrightarrow}\,
   \frac{Z^2}{4}|\psi(0,\hat{\bf x}_1)|^2+|\nabla_1\varphi_1(0,\hat{\bf x}_1)|^2
   -Z\psi(0,\hat{\bf
   x}_1)\big(\omega\cdot\nabla_1\varphi_1(0,\hat{\bf x}_1)\big).
   \nonumber
\end{align}
Using again \(\int_{\mathbb{S}^2}\omega\,d\omega=0\), it follows from
\eqref{eq:t1}--\eqref{eq:secondTermT1}, and Lebesgue's Theorem of
Dominated Convergence that 
\begin{align}
  \label{eq:limitT}\nonumber
  \lim_{r\downarrow0}\widetilde t_1(r)&=  
  \frac{Z^2}{4}\widetilde\rho_1(0)
  +\int_{\mathbb{S}^2}\int_{\R^{3N-3}}|\nabla_1\varphi_1(0,\hat{\bf x}_1)|^2
  \,d\hat{\bf x}_1 d\omega\\
  &\quad +\int_{\mathbb{S}^2}\int_{\R^{3N-3}}|\nabla_{\hat{\bf x}_1}\psi(0,\hat{\bf x}_1)|^2
  \,d\hat{\bf x}_1 d\omega\,.
\end{align}
This proves \eqref{eq:t(0)}.
Combining this with \eqref{eq:t'-a} (using \eqref{eq:partKin}) proves
\eqref{eq:tCusp}. 
\subsection*{\bf Dominant.}
We turn to finding a dominant to the integrand in \eqref{eq:der1}. We
shall apply results from \cite{CMP2}, recalled in
Appendix~\ref{app:B}. 

From  the {\it a priori} estimate \eqref{apriori} in
Theorem~\ref{thm:main:apriori} and \eqref{eq:der1} follows that, for
almost all \((x,\hat{\bf x}_1)={\bf x}\in\mathbb R^{3N}\) (choose,
e.g., \(R=1, R'=2\))
\begin{align}
  \label{eq:splitting}
  \Big|
     \omega\cdot\nabla_{1}\big(|\nabla\psi|^2\big)&(x,\hat{\bf x}_1)
  \Big|
  \leq C\, \big|\nabla\psi(x,\hat{\bf
    x}_1)\big|\,\|\psi\|_{L^\infty(B_{3N}((x,\hat{\bf x}_1),2))}\\  
  &+
  \Big|\Big(\sum_{\ell=1}^{N}\sum_{k=1}^{3}\sum_{m=1}^{3}
  2\frac{x_{1,k}}{|x_1|}\Big(\frac{\partial\psi}{\partial 
  x_{\ell,m}}\Big)\psi\frac{\partial^2 F_{\text{\rm cut}}}{\partial
  x_{1,k}\partial 
  x_{\ell,m}}\Big)(x,\hat{\bf x}_1)\Big|\,, \nonumber
\end{align}
with \(F_{\text{\rm cut}}=F_{2,\text{\rm cut}}+F_{3,\text{\rm cut}}\)
as in Definition~\ref{Cutoff}. 

Now, using the exponential decay of \(\psi\)
\eqref{eq:exp-dec} (assumed), and of \(\nabla\psi\) \eqref{eq:dec_grad_psi},
we get that there exist constants \(C,\gamma>0\) such that
\begin{align}
  \label{eq:firstDominant}
  \big|\nabla\psi(x,\hat{\bf
    x}_1)\big|\,\|\psi\|_{L^\infty(B_{3N}((x,\hat{\bf
    x}_1),2))}\leq
  Ce^{-\gamma|\hat{\bf x}_1|}
  \quad \text{for almost all } x\in\mathbb R^{3}.
\end{align}
This provides a dominant (in the sense of \eqref{eq:dominant}) for the
first term in \eqref{eq:splitting}.

We need to find a dominant for the second term in
\eqref{eq:splitting}. First recall that \(F_{\text{\rm
    cut}}=F_{2,\text{\rm cut}}+F_{3,\text{\rm cut}}\). With \(F_2\) as
in \eqref{F2} we have \(F_{2,\text{\rm cut}}=F_2+(F_{2,\text{\rm
    cut}}-F_2)\), with 
\begin{align}
  \label{eq:diffF-2s}
  (F_{2,\text{\rm cut}}-F_2)({\bf x})
  &=\sum_{i=1}^N{}-\frac{Z}{2}\,
  \big(\chi(|x_{i}|)-1\big)\,|x_{i}|     
  \\&\qquad\qquad\qquad
  +\sum_{1\le i<j\le
    N}\frac{1}{4}\,\big(\chi(|x_i-x_j|)-1\big)\,|x_i-x_j|\,.\nonumber 
\end{align}
Note that \(\partial^2\big[(\chi(|x|)-1)|x|\big]\) is bounded in
\(\R^3\) for all second derivatives \(\partial^2\), due to the
definition \eqref{eq:def_cutoff} of \(\chi\). Using the exponential
decay of \(\psi\) 
\eqref{eq:exp-dec}, and \(\nabla\psi\) \eqref{eq:dec_grad_psi}, we
therefore get a dominant for the term
\begin{align*}
  \Big(\sum_{\ell=1}^{N}\sum_{k=1}^{3}\sum_{m=1}^{3}
   2\frac{x_{1,k}}{|x_1|}\Big(\frac{\partial\psi}{\partial 
    x_{\ell,m}}\Big)\psi\frac{\partial^2 (F_{2,\text{\rm cut}}-F_2)}{\partial
    x_{1,k}\partial 
    x_{\ell,m}}\Big)(x,\hat{\bf x}_1)\,.
\end{align*}
A tedious, but straightforward computation gives that
\begin{align}
  \label{eq:finalSecondDerF2}
  &\sum_{\ell=1}^{N}\sum_{k=1}^{3}\sum_{m=1}^{3}2\frac{x_{1,k}}{|x_1|}
  \Big(\frac{\partial\psi}{\partial x_{\ell,m}}\Big) \psi
  \frac{\partial^2 F_{2}}{\partial 
  x_{1,k}\partial x_{\ell,m}}\\
  &=\frac{1}{2}\psi\sum_{i=2}^{N}
  \left[\frac{1}{|x_1-x_i|}\frac{x_1}{|x_1|}\cdot
  \big(\nabla_1\psi-\nabla_i\psi\big)\right.\nonumber\\  
  &\qquad\qquad\qquad\qquad\left.-\Big(\frac{x_1}{|x_1|}\cdot
  \frac{x_1-x_i}{|x_1-x_i|^3}\Big)
  \Big[\big(\nabla_1\psi-\nabla_i\psi\big)\cdot(x_1-x_i)\Big]\right]\,.\nonumber
\end{align}
We first remark that, again using exponential decay of \(\psi\) and
\(\nabla\psi\), we get the estimate
\begin{align}
  \label{eq:firstF2}
  \Big|\Big(\frac{1}{2}\psi\sum_{i=2}^{N}
  \frac{1}{|x_1-x_i|}\frac{x_1}{|x_1|}&\cdot
  \big(\nabla_1\psi-\nabla_i\psi\big)\Big)(r\omega,\hat{\bf x}_1)\Big|
  \\&\le C
  \sum_{i=2}^{N}\Big[\frac{e^{-c|x_i|}}{\dist(L(\omega),x_i)}\,
  \Big(\prod_{j=2,j\neq i}^{N}e^{-c|x_j|}\Big)\Big]\,,
  \nonumber
\end{align}
where \(\dist(L(\omega),y)\) is the distance from \(y\in\R^3\) to the
line \(L(\omega)\) spanned by \(\omega\in\mathbb{S}^2\). Note that for
fixed \(\omega\in\mathbb{S}^2\), the function
\(e^{-c|y|}/\dist(L(\omega),y)\) is integrable in \(\R^3\), and its
integral is independent of \(\omega\in\mathbb{S}^2\). Therefore the
right side of \eqref{eq:firstF2} is integrable in \(\R^{3N-3}\) for
fixed \(\omega\in\mathbb{S}^2\), and   is integrable in
\(\R^{3N-3}\times\mathbb{S}^2\) (all uniformly for \(r\in\R\)). 

The argument is similar for the second term in
\eqref{eq:finalSecondDerF2}. 

We are left with considering the terms in \eqref{eq:splitting} with 
\(F_{3,\text{\rm cut}}\). Let
\begin{align*}
   f_3(x,y)&=C_0Z(x\cdot y)\ln\!\big(|x|^2+|y|^2\big)
\end{align*}
(so that \( F_3({\bf x})=\sum_{i<j}f_3(x_i,x_j)\), see \eqref{F3}). 
For all second derivatives \(\partial^2\) we easily get, due to the
definition \eqref{eq:def_cutoff} of \(\chi\), 
\begin{align*}
  \partial^2\big[\chi(|x|)&\chi(|y|)f_3(x,y)\big]
  =\chi(|x|)\chi(|y|)\partial^2f_3(x,y) + g_3(x,y)\\
  &=C_0Z\chi(|x|)\chi(|y|)\ln\!\big(|x|^2+|y|^2\big)\partial^2(x\cdot y)
  +\widetilde{g}_3(x,y)\,,
\end{align*}
with \(g_3, \widetilde{g}_3\) bounded on \(\R^6\). Therefore,
defining, for all second derivatives \(\partial^2\) and with \(\chi\)
as above,
\begin{align}
  \label{eq:F3New}
  \widetilde{{\partial^2F}}_{3,\text{\rm cut}}({\bf x})
  :=C_0Z\!\!\!\sum_{1\le i<j\le N}
  \!\!\!\chi(|x_i|)\chi(|x_j|)\ln\!\big(|x|^2+|y|^2\big)\partial^2(x\cdot y)\,, 
\end{align}
we get a dominant for the term
\begin{align*}
  \Big[\sum_{\ell=1}^{N}\sum_{k=1}^{3}\sum_{m=1}^{3}
  2\frac{x_{1,k}}{|x_1|}
  \Big(\frac{\partial\psi}{\partial x_{\ell,m}}\Big)
  \psi\Big(
  \frac{\partial^2F_{3,\text{\rm cut}}}{\partial x_{1,k}\partial x_{\ell,m}}
  -\frac{\widetilde{{\partial^2F}}_{3,\text{\rm cut}}}{\partial
  x_{1,k}\partial x_{\ell,m}} 
\Big)\Big](x,\hat{\bf x}_1)\,,
\end{align*}
using the exponential decay of \(\psi\) and \(\nabla\psi\). We find
that 
\begin{align}
  \label{eq:finalSecondDerF3}
  \sum_{\ell=1}^{N}\sum_{k=1}^{3}\sum_{m=1}^{3}&2\frac{x_{1,k}}{|x_1|}
  \Big(\frac{\partial\psi}{\partial x_{\ell,m}}\Big) \psi
  \frac{\widetilde{{\partial^2 F}}_{3,\text{\rm cut}}}{\partial 
  x_{1,k}\partial x_{\ell,m}}\\
  &=2C_0Z\psi\sum_{i=2}^{N}\chi(|x_1|)\chi(|x_i|)
  \Big[\ln(|x_1|^2+|x_i|^2)\Big(\frac{x_1}{|x_1|}\cdot\nabla_i\psi\Big)\Big]\,. 
  \nonumber
\end{align}
Note that, by the definition of \(\chi\), 
\begin{align*}
  \chi(|x|)\chi(|y|)\,\big|\ln(|x|^2+|y|^2)\big|
  \le \chi(|y|)\big[ |\ln(|y|^2)|+3\big]\,,
\end{align*}
and so, again by the exponential decay of \(\psi\) and \(\nabla\psi\),
\begin{align}
  \label{eq:secondF3cutBound}\nonumber
  \Big|2C_0Z\psi&\sum_{i=2}^{N}
  \Big[\chi(|x_1|)\chi(|x_i|)\ln(|x_1|^2+|x_i|^2)\Big(\frac{x_1}{|x_1|}
  \cdot\nabla_i\psi\Big)\Big](r\omega,\hat{\bf x}_1)\Big|
  \\&\le
  C\Big(\sum_{i=2}^{N}\chi(|x_i|)\big[|\ln(|x_i|^2)|+3\big]\Big)
  \Big(\prod_{j=2}^{N}e^{-c|x_j|}\Big)
\end{align}
for all \(\omega\in\mathbb{S}, r\in\R\) and (almost) all \(\hat{\bf
  x}_1\in\mathcal{S}_1(\omega)\). This provides a dominant for the
term 
\begin{align*}
  \Big[\sum_{\ell=1}^{N}\sum_{k=1}^{3}\sum_{m=1}^{3}
  2\frac{x_{1,k}}{|x_1|}
  \Big(\frac{\partial\psi}{\partial x_{\ell,m}}\Big)
  \psi\frac{\widetilde{\partial^2F}_{3,\text{\rm cut}}}{\partial
  x_{1,k}\partial x_{\ell,m}} 
  \Big)\Big](x,\hat{\bf x}_1)\,,
\end{align*}
and we have therefore provided a dominant, in the sense of of
\eqref{eq:dominant}, for the integrand in \eqref{eq:der1}. 

This finishes the proof of Proposition~\ref{prop:t'}.
\qed
\appendix
\section{Two useful lemmas}\label{app:lem}
\label{app:A}
The following lemma is Lemma~2.9 in \cite{CMP2}; we include it, without
proof, for the convenience of the reader. (The proof is simple, and
can be found in \cite{CMP2}). 
\begin{lemma}\label{usefullemma}
\label{lem:XdotG}
Let \(G:U\to\mathbb R^{n}\) for \(U\subset\mathbb R^{n+m}\) a
neighbourhood of a point \((0,y_{0})\in\mathbb R^{n}\times\mathbb
R^{m}\). Assume \(G(0,y)=0\) for all \(y\) such that \((0,y)\in
U\). Let
\begin{align*}
  f(x,y)=\left\{\begin{array}{cc}
               \frac{x}{|x|}\cdot G(x,y)& x\neq 0, \\
               0& x=0. \\
  \end{array}\right. 
\end{align*}
Then, for \(\alpha\in(0,1]\), 
\begin{align}
  \label{eq:lem_G=0}
  G\in C^{0,\alpha}(U;\mathbb R^{n})\Rightarrow f\in C^{0,\alpha}(U).
\end{align}
Furthermore, $\| f \|_{C^{\alpha}(U)} \leq 2\| G \|_{C^{\alpha}(U)}$.
\end{lemma}

The following lemma is used to evaluate certain integrals.
\begin{lemma}\label{eq:dotprod}
Let \(a,b\in\mathbb R^3\) and let \(A\in M_{3\times3}(\C)\).

Then
\begin{align}
  \label{eq:intOmega}
  \int_{\mathbb S^2}(\omega\cdot a)(\omega\cdot b)\,d\omega
  &=\frac{4\pi}{3}(a\cdot b)
  =\frac13(a\cdot b)\int_{\mathbb S^2}\,d\omega\,,
  \\\label{eq:intSymMatrix}
  \int_{\mathbb{S}^2} \omega\cdot(A\omega)\,d\omega
  &=\frac{4\pi}{3}\Tr(A)
  =\frac13\Tr(A)\int_{\mathbb S^2}\,d\omega\,.
\end{align}
\end{lemma}
\begin{proof}
Both \eqref{eq:intOmega} and \eqref{eq:intSymMatrix} follow from the identity
\begin{align}
  \label{eq:star}
  \int_{{\mathbb S}^2} \omega_i \omega_j \,d\omega =
  \frac{4\pi}{3}\,\delta_{i,j}\,, 
\end{align}
which we now prove.

For $i \neq j$, the integrand $ \omega_i \omega_j$ is odd as a
function of $\omega_j$, which implies \eqref{eq:star} in that case. 
For $i=j$ we calculate, using rotational symmetry,
$$
  \int_{{\mathbb S}^2} \omega_i^2  \,d\omega =
  \frac{1}{3} \int _{{\mathbb S}^2} (\omega_1^2 + \omega_2^2  +
  \omega_3^2)  \,d\omega = \frac{4\pi}{3}. 
$$
This finishes the proof of \eqref{eq:star} and by consequence of
Lemma~\ref{eq:dotprod}. 
\end{proof}
\section{Regularity of the eigenfunction \(\psi\)}
\label{app:B}
The following two theorems were proved in \cite{CMP2}.
\begin{thm} {\rm (}\cite[Theorem~1.1 for atoms]{CMP2}{\rm)}\label{thm:main:Jastrow}
Suppose $\psi$ is a solution to 
$H\psi=E\psi$ in \(\Omega\subseteq\R^{3N}\) where $H$ is given by \eqref{H}. 
Let \(\mathcal F=e^{F_2+F_3}\) with
\begin{align}
 \label{F2}
  &F_2({\mathbf
  x})=\sum_{i=1}^N{}-\frac{Z}{2}|x_{i}| 
  +
  \sum_{1\le i<j\le N}\frac{1}{4}\,|x_i-x_j|,\\
  \label{F3}
  &F_3({\mathbf x})=
  C_0\sum_{1\le i<j\le N}Z\,
  (x_{i}\cdot x_{j})\,\ln(|x_{i}|^2+
  |x_{j}|^2),
\end{align}
where $C_0=\frac{2-\pi}{12\pi}$.
Then  \( \psi=\mathcal F\phi_3\) with \( \phi_3\in C^{1,1}(\Omega)\).
\end{thm}
\begin{defn}\label{Cutoff}
Let \(\chi\in C_{0}^{\infty}(\R)\), \(0\leq \chi\leq1\), with
\begin{align}
  \label{eq:def_cutoff}
  \chi(x)=
  \begin{cases}
    1& \text{ for } |x|\leq1 \\
    0& \text{ for } |x|\geq2.
  \end{cases}
\end{align}
We define
\begin{equation}\label{F23cut}
  F_{\text{\rm cut}}=F_{2,\text{\rm cut}}+F_{3,\text{\rm cut}},
\end{equation}
where 
\begin{align}
 \label{eq:def_F2cut}
  &F_{2,\text{\rm cut}}({\mathbf
  x})=\sum_{i=1}^N{}-\frac{Z}{2}\,
  \chi(|x_{i}|)\,|x_{i}|     
  +\frac14\!
  \sum_{1\le i<j\le N}\!\!\!\!\!
  \chi(|x_i-x_j|)\,|x_i-x_j|,
  \\
 \label{eq:def_F3cut}
  &F_{3,\text{\rm cut}}({\mathbf x})=
  C_0\sum_{1\le i<j\le N}Z\,
  \chi(|x_{i}|)\chi(|x_{j}|)
  ( x_{i}\cdot x_{j})\,\ln(|x_{i}|^2+
  |x_{j}|^2),
\end{align}
and where $C_0$ is the constant from \eqref{F3}.
We also introduce $\phi_{3,\text{\rm cut}}$ by
\begin{equation}\label{phcut}
  \psi =e^{F_{\rm cut}}\phi_{3,\text{\rm cut}}.
\end{equation}
\end{defn}
\begin{thm}{\rm (}\cite[Theorem~1.5 for
  atoms]{CMP2}{\rm)}\label{thm:main:apriori} 
Suppose $\psi$ is a  solution to $H\psi=E\psi$ in \(\R^{3N}\). 
Then for all \(0<R<R'\) there exists a constant $C(R,R')$, not
depending on $\psi$ nor \(\mathbf x_0\in\R^{3N}\), such that for any
second order derivative, 
\begin{equation*}
  \partial^2=\frac{\partial^2}
  {\partial x_{i,k}\partial x_{j,\ell}},\:\:i,j=1,2,\dots, N, \:\:\: 
  k,\ell =1,2,3,
\end{equation*}
the following {\emph{a priori}} estimate holds:
\begin{equation}\label{apriori}
  \|\partial^2\psi -\psi\,\partial^2\!
  F_{\text{\rm cut}}\|_{L^\infty(B_{3N}(\mathbf x_0,R))}\le
  C(R,R')\|\psi\|_{L^\infty(B_{3N}(\mathbf x_0,R'))}.
\end{equation}
\end{thm}

\begin{acknowledgement}
Parts of this work have been carried out at various
institutions, whose hospitality is gratefully acknowledged:
Mathematisches Forschungsinstitut Oberwolfach 
(SF, T\O S), Erwin Schr\"{o}\-dinger Institute (SF, T\O
S), Universit\'{e} Paris-Sud (T\O S), and the IH\'ES (T\O S).
Financial support from the 
European Science Foundation Programme {\it Spectral Theory and Partial 
  Differential Equations} (SPECT), and EU IHP network 
{\it Postdoctoral Training Program in Mathematical Analysis of
Large Quantum Systems},
contract no.\
HPRN-CT-2002-00277, is
gratefully acknowledged.
T\O S was partially supported
by the embedding grant from The Danish National
Research Foundation: Network in Mathematical Physics and Stochastics, and
by the European Commission through its 6th Framework Programme
{\it Structuring the European Research Area} and the contract Nr.
RITA-CT-2004-505493 for the provision of Transnational Access
implemented as Specific Support Action.
\end{acknowledgement}
\providecommand{\bysame}{\leavevmode\hbox to3em{\hrulefill}\thinspace}
\providecommand{\MR}{\relax\ifhmode\unskip\space\fi MR }
\providecommand{\MRhref}[2]{%
  \href{http://www.ams.org/mathscinet-getitem?mr=#1}{#2}
}
\providecommand{\href}[2]{#2}

\end{document}